Imperial College of Science, Technology and Medicine
Department of Mathematics

# The dynamics of the Koopman-van Hove angular and linear momentum operators

Remy Messadene

**This is my unaided work unless stated otherwise.**

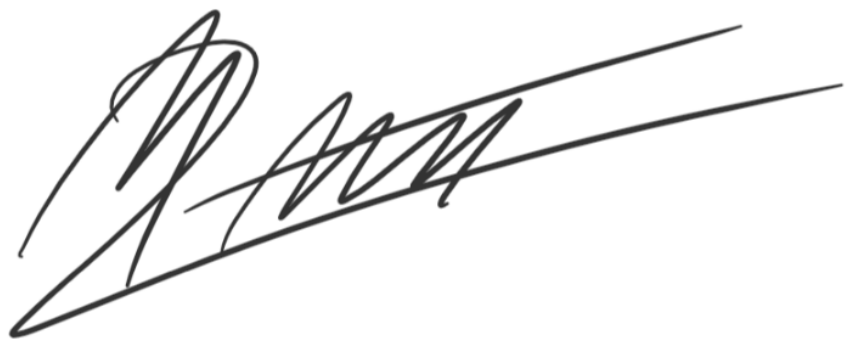

## Abstract

Recently, Tronci and Gay-Balmaz [1] solved the problem of expressing the classical-quantum coupling in a Hamiltonian theory by proposing a new formalism for probabilistic classical mechanics: the Koopman-van Hove formulation. First, we rigourously expose this new framework before presenting a construction for the operatorial analogues to the classical angular and linear momentum operators. We then investigate the group actions generating their average as momentum maps as well as their associated dynamics. Finally, we apply the Koopman-van Hove formulation to the Kepler problem, harmonic and anharmonic oscillators, highlight new terms appearing in the dynamics of the two above operators.

"If I have seen further, it is by standing on the shoulders of giants."

*Isaac Newton*

# Contents

# Chapter 1

# Introduction

We will expose the fundamentum of a new formulation of probabilistic classical mechanics, namely the Koopman-van Hove formulation [1]. First, we construct the analogues of the linear and angular momentum operators, extract their conservation laws as well as the group action generating their average as momentum map. We then conclude by the application of this formalism to the Kepler problem, the harmonic, and anharmonic oscillators.

The Kepler problem has been intimately linked to the birth and development of many fundamental mechanical theories. Indeed, the understanding of the motions of celestial bodies has been intriguing for centuries. In 1609 and 1619, out of the accurate celestial measurements of Brahe, Kepler extrapolated 3 empirical laws [2], the celebrated Kepler's laws. The third one in particular led Newton to his law of gravitation [3], before he established the fundamental equations of motion of mechanics. In addition to motivating Newton's gravitation law, the problem of determining the motion of two celestial bodies (which is now referred as the Kepler's problem) played an important role in establishing Newton's equations $\mathbf{F} = m\mathbf{a}$ as the building block of classical mechanics. Since then, the Kepler problem has played a key role in the development of fundamental concepts in the determination of symmetries in classical mechanics. For example, it was used by Smale [4] to expose and motivate his concept of special functions that are now called momentum maps and by Jacobi [5] in his elimination of the node. The Kepler problem also had a key role in the work of Euler, Lagrange, Laplace, Runge, Lenz, Poisson, Jacobi, Hamilton, Lie, Noether, Arnold and many others [6, 7, 8, 9].

On the other hand, the Kepler problem has also been of fundamental importance during the emergence of quantum theory. For instance, in 1885, physicists were aware of the existence of emission spectrum lines but were lacking the tool to accurately predict them. By using tables of empirical data, Balmer solved the problem in the case of the hydrogen atom and determined a formula to predict its emission spectrum lines. Although inspiring, Balmer's formula did not give any explanation concerning the underlying mechanism. The first model to explain the repartition of the emission spectrum lines of the Hydrogen atom is the Bohr-Rutherford model, the quantum analogue of the Kepler problem modelling the electron as a classical object experiencing a central electromagnetic force and rotating around a positively charged nucleus. After 20 years of refinement, correction and development, Schrödinger was the first to give a fully detailed explanation of the spectral emission in the then-Bohr-Sommerfeld model.

Many fundamental questions emerged after the discovery of quantum mechanics, *e.g.* What are the links between classical and quantum mechanics? Could one construct a dynamical theory for a generalized density operator that takes into account the interactions between objects of classical and quantum nature? In order to study the interactions and parallels between macroscopic and quantum systems, physicists initially tried to augment the classical theory by adding quantum effects but this formulation, now called the 'old quantum theory', has never been complete or self-consistent. The formulation of quantum mechanics proposed by Schrödinger was then exploited by Koopman and Von-Neumann who attempted the converse, namely constructing an operatorial formulation of classical mechanics. Unfortunately, although natural and elegant, its lack of Hamiltonian structure made it limited.

Over the last four decades, several streams of work attempted to solve the problem of describing the classical-quantum coupling with a Hamiltonian theory (for example, some people proposed the analogue of a classical-quantum Liouville equation [10] and attempted to equip it with a Hamiltonian structure [11, 12, 13], until the proof of its non-feasibility [12]). Recently, following Sudarshan's idea [14] to couple classical and quantum dynamics by reconsidering the Koopman-von Neumann formulation of classical mechanics, Tronci and Gay-Balmaz [1] proposed instead an alternative version of it, the Koopman-van Hove one, where the Hamiltonian function coincides with the physical energy. The latter was then exploited to derive a Hamiltonian theory for the classical-quantum coupling.

This Koopman-van Hove theory is a new way of expressing classical dynamics, and its study might give us a deeper understanding of symmetries in probabilistic classical mechanics. The formalism might allow the extension of many standard techniques of quantum mechanics to classical problems and may facilitate the resolution of classical non-linear problems by linearising them.

Motivated by the above reasons, we consider the Kepler problem, but also another fundamental model in physics, the harmonic and anharmonic oscillators, to illustrate the Koopman-van Hove formulation of classical mechanics [1].

We will proceed in the following manner. First, **Chapter 2** acts as a gentle introduction to geometric mechanics and presents some fundamental tools that will be used later on in this report. **Chapter 3** rigorously exposes the mathematical framework of the classical Koopman-von Neumann theory, its limitations and then the construction of the Koopman-van Hove formulation proposed in [1]. We further develop this recent theory in **Chapter 4** by presenting the construction of the two operators that are analogue to the classical linear and angular momentum. We also investigate the dynamics of these operators before determining the group action generating their average as momentum maps. To conclude, we illustrate these new objects in the special case of the Kepler problem in **Chapter 5**, and the (an)harmonic oscillators in **Chapter 6**; the Koopman-van Hove dynamics will be compared with its Koopman-von Neumann analogue and the new contributions in the dynamics of the two above operators will be highlighted.

# Chapter 2

# Preliminaries: elements of geometric mechanics

Geometric mechanics [15, 16, 17, 18] deals with systems whose dynamics can be reformulated as variational problems. This framework allows the determination of their constants of motion as well as the reduction of their complexity (*i.e.* the reduction of its number of degrees of freedom).

In this section, we will provide an introduction to geometric mechanics and expose a selection of tools that will be used later on during our study. An experienced reader may wish to skip this chapter.

## 2.1 The variational problem and its reformulation

Let $M$ be a differentiable manifold. Consider a Lagrangian function $L : TM \to \mathbb{R}$, where $TM$ denotes the tangent bundle associated with $M$.

The core problem one is interested in solving is the determination of a path in M among a smoothly parametrized family of curves which locally minimizes the action functional associated with $L$, i.e. the determination of a path $\gamma$ such that

$$\boxed{\delta \int_a^b L(\gamma_\epsilon(t), \dot{\gamma}_\epsilon(t))dt := \frac{d}{d\epsilon}\Big|_{\epsilon=0} \int_a^b L(\gamma_\epsilon(t), \dot{\gamma}_\epsilon(t))dt = 0} \tag{2.1}$$

where $\gamma : [\alpha; \beta] \times [a, b] \to M$ defined as $\gamma(\epsilon, t) := \gamma_\epsilon(t)$ is the smooth variation of $\gamma$. Equivalently, $\gamma$ is a solution of the above variational problem (2.1) if and only if it satisfies the Euler-Lagrange equations:

$$\frac{d}{dt}\frac{\partial L}{\partial \dot{q}} - \frac{\partial L}{\partial q} = 0. \tag{2.2}$$

Now, as the PDE (2.2) can become very complex to solve for the variables $(q, \dot{q})$ in many cases, one would want to express it in an different form which might be easier to solve. To achieve this goal, one possible reformulation exploits the notion of Legendre transform. Assuming that a Lagrangian $L$ is smooth enough, the problem can be reformulated in terms of two coupled ODEs, namely Hamilton's equations.

**Definition 2.1** (Legendre transformation)**.** *Consider a $C^1$-Lagrangian $L$. The **Legendre transformation** $\mathcal{LT}[L] : TM \to T^*M$ is defined as*

$$\mathcal{LT}[L](v_q)(w_q) \equiv \langle \mathcal{LT}[L](v_q), w_q \rangle := \frac{d}{ds}\Big|_{s=0} L(v_q + s \cdot w_q), \tag{2.3}$$

*where $v_q, w_q \in T_qM$, and the first identification comes from Riesz' representation Theorem. Moreover, $\mathcal{LT} : \{f : TM \to \mathbb{R} \mid f \in C^1\} \to \{f : T^*M \to \mathbb{R} \mid f \in C^1\}$.*

Consider a hyperregular Lagrangian L, *i.e. the Legendre transform of L is a diffeomorphism.* We define the Hamiltonian function $H : T^*M \to \mathbb{R}$ as the composition of the energy function $E$ with the inverse of the Legendre transform $\mathcal{LT}^{-1}[L]$ of $L$.

**Definition 2.2** (Hamiltonian and energy functions)**.** ***The Hamiltonian function*** *$H : T^*M \to \mathbb{R}$ associated with a hyperregular Lagrangian $L$ is defined as $H := E \circ \mathcal{LT}^{-1}$ where the **energy function** $E : TM \to \mathbb{R}$ is given as $E(v) := \langle \mathcal{LT}[L](v), v \rangle - L$.*

Finally, by using (2.1), the Euler-Lagrange equations transform into the **Hamilton's canonical equations**:

$$\begin{cases} \frac{\partial H}{\partial p}(q, p) = \dot{q} \\ \frac{\partial H}{\partial q}(q, p) = -\dot{p} \end{cases} \tag{2.4}$$

**Remark 2.1.** *For hyperregular $L$, one can freely transform from one formalism the other.*

**Theorem 2.1** (Euler-Lagrange and Hamilton's equations)**.** *Consider a hyperregular Lagrangian* $L : TM \to \mathbb{R}$*.* $\gamma \equiv \gamma(t)$ *is a solution of the Euler-Lagrange equations if and only if* $(q(t), p(t)) := \left(\gamma(t), \frac{\partial L}{\partial \dot{\gamma}}(\gamma(t), \dot{\gamma}(t))\right)$ *is a solution of Hamilton's equations* (2.4)*.*

*Proof.* Use the fact that the Lagrangian is hyperregular to invert the Legendre transform. Show then that both Ansätze satisfy Hamilton's equations and Euler-Lagrange's equations respectively. An explicit computation can be found in [16]. □

## 2.2 The reduction theory

### 2.2.1 The reduction

Many interesting geometrical features appear when one considers the transitive action of a connected Lie group $G$ on $M$ leaving $L$ invariant, i.e.

$$L(g.q, g.v) = L(q, v), \tag{2.5}$$

where $(q, v) \in TM$ and $g \in G$. Note that without restriction of the generality and for the purpose of this report, we restrict ourselves to left actions.

Start by fix initial conditions $(q_0, v_0) \in TM$. Then, by transitivity, for all $(q, v) \in TM$, there exists $t \in [a, b]$ and $\gamma \in C^\infty([a, b],\ M)$ such that $(\gamma(t), \dot{\gamma}(t)) = (q, v) \in TM$. Moreover, there exists $\{g_s\}_{s\in[a,b]} \subseteq G$ such that

$$\gamma(t) = g_t q_0 \; \forall t \in [a, b]. \tag{2.6}$$

Then, by using (2.5) and (2.6), a reduced Lagrangian can be derived.

**Definition 2.3** (Reduced Lagrangian)**.** *The* ***reduced Lagrangian*** $l : \mathfrak{g} \to \mathbb{R}$ ***associated with*** $\boldsymbol{L}$ *is defined as* $L(g.q, g.v) = L(g_t q_0, \dot{g}_t q_0) = L(q_0, g_t^{-1}\dot{g}_t q_0) =: l(\xi)$*, where* $\xi := g_t^{-1}\dot{g}_t \in \mathfrak{g}$ *and* $\mathfrak{g} := T_eG$ *denotes the Lie algebra associated with* $G$*.*

Without going into details, Hamilton's principle $\delta \int_a^b l(\xi)dt = 0$ yields a modified version of the Euler-Lagrange equations, called the **Euler-Poincaré equations**:

$$\frac{d}{dt}\frac{\delta l}{\delta \xi} = \mathrm{ad}^*_\xi \frac{\delta l}{\delta \xi}. \tag{2.7}$$

Analogously to the standard case, one defines the **reduced Hamiltonian** $h : \mathfrak{g}^* \to \mathbb{R}$ as the composition of a reduced energy function with the inverse of the reduced Legendre transform of $l$. More details concerning the Euler-Poincaré equations and reduced Hamiltonians can be found in [16].

### 2.2.2 Constant of motions and momentum maps

We are now interested in the determination of constant of motions, *i.e. quantities having a vanishing time derivative for all times.* First off, Emmy Noether's theorem [19] gives us a way of obtaining specific constants of motion in a particular setting. More precisely,

**Theorem 2.2** (Noether's theorem). *Let $q : [a, b] \to M$ be a path on $M$ and let $\Phi_{(\cdot)}((\cdot),(\cdot)) : \mathbb{R} \times M \times [a, b] \to M$ be a variation of $q$. Let $d\Phi_\epsilon(\cdot, t_0) : TM \to TM$ be the differential, for $\epsilon \in \mathbb{R}$ and $t_0 \in \mathbb{R}$ fixed. Consider now a Lagrangian $L : TM \to \mathbb{R}$ invariant under the action of the differential $d\Phi_\epsilon(\cdot, t_0)$ for all $\epsilon$, $t_0 \in \mathbb{R}$. Then, $F(t) := \frac{\partial L}{\partial \dot{q}}(q(t), \dot{q}(t)) \cdot \frac{\partial \Phi_s(q(t))}{\partial s}\Big|_{\epsilon=0}$ is a constant of motion, i.e. $\frac{d}{dt}F(t) = 0$ for all $t \in [a, b]$.*

The ideas of Noether's paper have been developped further by Smale, Weinstein, Marsden, Ratiu, Holm et al [4, 20, 21] and led to the concept of momentum mapping, which is intimately linked with the notions of symmetries and constant of motions.

Before defining what a momentum map is, we will recall the notions of Lie exponential map and infinitesimal generator.

**Definition 2.4** (Lie exponential map). *Let $\xi \in \mathfrak{g}$ and let $L_g : G \to G$ be the left action defined as $L_g(h) = g.h$ where "." denotes group multiplication and $g, h \in G$. Consider the vector field $X_\xi : G \to T_eG \equiv \mathfrak{g}$ defined as $X_\xi(g) := dL_g(e)(\xi)$ as well as the following ODE:*

$$\begin{cases} \frac{d}{dt}\big|_{t=0} g = X_\xi(g) \\ g(0) = Id \end{cases} \tag{2.8}$$

*where* $\{g_t\}_{t\in[\alpha,\beta]} \subset G$ *and* $[0,1] \subseteq [\alpha,\beta]$*. By the Picard-Lindelöf theorem, this ODE admits a solution* $g_\xi : [\alpha,\beta] \to M$ *depending on the initial condition* $\xi$*.*
***The Lie exponential map*** *is defined as*

$$\exp : \mathfrak{g} \to G, \ \exp(\xi) := g_\xi(1). \tag{2.9}$$

Before introducing the definition of the infinitesimal generator, let us state a useful lemma [22].

**Lemma 2.1** (Rescaling Lemma)**.** *Consider (2.8) and let* $g_\xi : [\alpha,\beta] \to G$ *be its solution. Then,*

$$exp(t\xi) := g_{t\xi}(1) = g_\xi(t) \tag{2.10}$$

*for all* $t$ *such that* $g_\xi(t)$ *is defined.*

*Proof.* Simply verify that $g_{t\xi}$ is a solution of the ODE system with initial conditions $\xi$ by taking into account the parametrization $\phi(\xi) := t\xi$. Conclude then by using the uniqueness statement of the Picard-Lindelöf theorem. □

**Definition 2.5** (Infinitesimal generator)**.** *Consider the left action of a Lie group* $G$ *on* $M$*. Choose* $\xi \in \mathfrak{g}$ *and consider the one-parameter subgroup* $\{\exp(\xi \cdot t)\}_{t\in\mathbb{R}} < G$*.*
*Then, the* ***infinitesimal generator*** $\xi_M$ *associated with* $\xi$ *is the vector field on* $M$ *defined as*

$$\xi_M(x) := \frac{d}{dt}\Big|_{t=0} \exp(t\xi)x \ \in T_{\exp(t\xi)}M. \tag{2.11}$$

**Remark 2.2.** *By using the rescaling lemma 2.1, the infinitesimal generator* $\xi_M$ *is interpreted as the velocity of the curve* $\gamma(t) := \exp(t\xi)x$ *starting at* $\gamma(0) = x$ *with initial velocity* $\xi$ *for a fixed* $x \in M$*.*

Now that these key objects have been defined, one defines a momentum map as follows.

**Definition 2.6** (Momentum map)**.** *Let* $(M, \{\cdot,\cdot\})$ *be a differentiable manifold equipped with a Poisson bracket. Let* $\Phi : G \times M \to M$ *be the left action of a Lie group* $G$ *on* $M$ *preserving the Poisson bracket (called canonical action), i.e.* $\{F \circ \Phi(g,\cdot), G \circ \Phi(g,\cdot)\} = \{F, G\}$ *for all* $F, G : C^1(M,\mathbb{R})$ *and for all* $g \in G$*. Then,* $J : M \to \mathfrak{g}^*$ *is called the* ***momentum map*** $J$ ***associated with the action*** $\Phi$ *if and only if* $X_{J_\xi} = \xi_M$ *where* $J_\xi(x) := \langle J(x), \xi\rangle$ *for all* $\xi \in \mathfrak{g}$*.*

Moreover, for particular group actions, momentum maps are constants of motion.

**Theorem 2.3** (Momentum maps and constants of motion). *If a Hamiltonian function $H : (M, \{\cdot,\cdot\}) \to \mathbb{R}$ is invariant under a canonical Lie group action $\Phi$, then the momentum map associated with $\Phi$ is a conserved quantity under the flow associated with the vector field $X_H$.*

*Proof.* This is theorem 11.4.1 in [18]. □

We conclude this chapter by stating an alternative characterization of momentum maps in the special case of symplectic vector spaces that we will use later on.

**Proposition 2.1** (Alternative characterization). *Let $V$ be a symplectic vector space and $\omega$ its associated symplectic two-form. Let $\Phi : G \times V \to V$ be a unitary action of a Lie group $G$ on $M$ preserving the Poisson bracket.Then, $J : V \to \mathfrak{g}^*$ is a momentum map if and only iff $\langle J(v), \xi\rangle = \frac{1}{2}\omega(\xi_V(v), v)$ for all $\xi \in \mathfrak{g}$.*

*Proof.* This is a special case of example (h) presented in chapter 11.4 in [18]. □

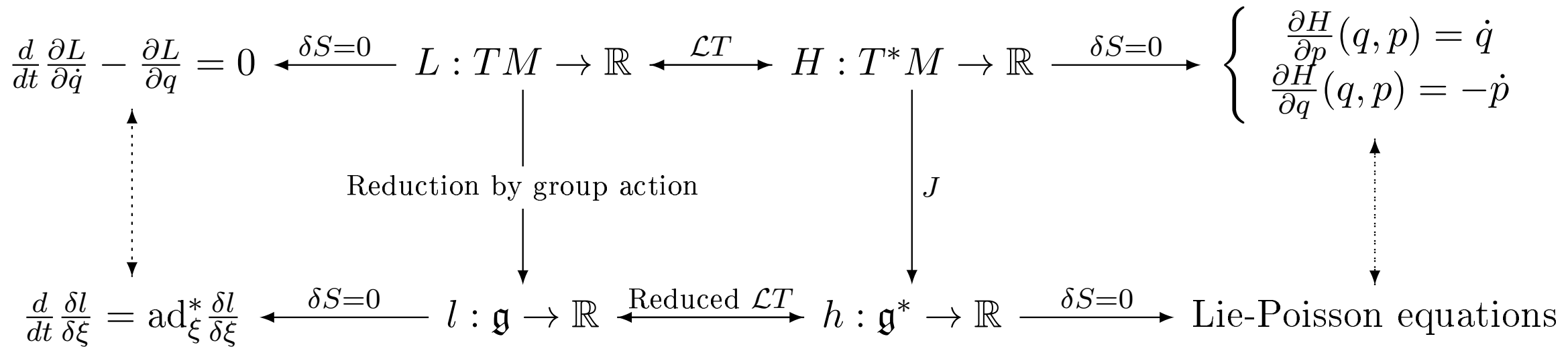

Figure 2.1: Diagram illustrating some key transformations in geometric mechanics where $\delta S = 0$ denotes the use of a variational principle. The dashed arrows stand for additional existing relations [18] that have not been mentioned in this chapter.

# Chapter 3

# The Koopman-van Hove formulation of classical mechanics

The Koopman-von Neumann formalism is a standard and well-known description of classical mechanics which is structurally similar to the quantum theory, *i.e.* built on an operatorial theory acting on classical mixed states which are elements of a Hilbert space. Because of its structural similarities with quantum mechanics, this formalism opens the door to the application of many quantum techniques to classical problems. However, while being an elegant theory attempting to provide a strong correspondence with quantum mechanics, the Koopman-von Neumann formalism has some very fundamental issues, that have been solved only very recently [1].

As a prelude, we start by providing insight about the meaning of pure and mixed classical states and explicitly show how deterministic and probabilistic mechanics relate to each other. Once this is done, the standard Koopman-Von Neumann formalism will be exposed, as well as its physical limitations. We conclude this chapter by presenting a recent alternative formulation, namely the Koopman-van Hove formulation [1]. We will extensively use [18] and [23].

## 3.1 Elements of probabilistic classical mechanics

A motion in classical mechanics can be described either deterministically or probabilistically, respectively by pure or mixed states. In other words, in the pure state description, the information concerning the position of the particles is known, exact, and is totally determined while in the mixed state one, one has no exact information about the position of the particles in the system; only expectation values can be extracted.

Let us now formally define what is meant by classical pure and mixed state.

**Definition 3.1** (Classical pure state)**.** ***A pure state*** *in classical mechanics is a deterministic differentiable path* $\gamma : [a,b] \to M$ *in a symplectic manifold* $(M,\ \omega)$*, parametrised by a time in* $[a,b] \subset \mathbb{R}$, $a \neq b$*, and satisfies Newton's equation* $m\ddot{\gamma}(t) = F_{tot}(\gamma(t))$*, where* $F_{tot}(\gamma(t))$ *is the total force acting on the particles.*

**Definition 3.2** (Classical mixed state)**.** ***A mixed state*** *in classical mechanics is a smooth probability density function* $\rho : [a,b] \times M \to [0,1]$ *normalized over the phase space, i.e.* $\int_M \rho(t_0, x)dx = 1$*,s for all* $t_0 \in [a,b]$ *satisfying Liouville's equation*

$$\frac{\partial \rho}{\partial t} = \{H, \rho\}, \tag{3.1}$$

*where* $H : M \to \mathbb{R}$ *is the Hamiltonian (or energy function) of the system,* $\{\cdot,\cdot\}$ *is the* ***Poisson bracket associated with the symplectic form*** $\omega$*, i.e.*

$$\{F, G\} := \omega(X_F, X_G) \tag{3.2}$$

*for* $F, G : M \to \mathbb{R}$ *where the* ***Hamiltonian vector field*** $X_F$ *associated with* $F$ *is uniquely determined by* $\omega(X_F, \cdot) = dF(\cdot)$.

**Remark 3.1** (Hamiltonian vector field for a symplectic vector space)**.** *For symplectic vector space equipped with the canonical symplectic form* $(\mathbb{R}^{2n}, \mathbb{J})$*, the Hamiltonian vector field* $X_F$*, for* $F : \mathbb{R}^{2n} \to \mathbb{R}$ *has the following form:*

$$\begin{aligned}
\nabla F(\cdot) \equiv dF(\cdot) &= \omega(X_F, \cdot) && \\
&= -\omega(\cdot, X_F) && \textit{antisymmetry} \\
&= -(\cdot)\,\mathbb{J}X_F && \textit{by definition of } \omega \\
\iff X_F &= \mathbb{J}\nabla F(\cdot) && \textit{as } \mathbb{J}^2 = -\mathbb{J}
\end{aligned} \tag{3.3}$$

Before ending this section and investigating further a mixed state description of classical me-chanics, one should ask themself how such a description would relate to the traditional pure state one. Even though the link might be clear, it remains fundamental to show that the mixed state representation of classical mechanics physically encompasses the pure state one. In other words, classical mechanics can be expressed in terms of mixed states.

**Proposition 3.1** ("Liouville implies Newton"). *Consider $H : T^*M \to \mathbb{R}$ hyperregular. Then Newton's equations can be recovered from Liouville's equation.*

*Sketch of proof.* Consider a solution of Liouville equation of the form $\rho_t(z) := \delta(z - \tilde{z}(z_0, t))$ where $\tilde{z} : \mathbb{R}^6 \times [a, b] \to \mathbb{R}^6$ is a smooth function. Without restriction of the generality, assume that $\tilde{z}$ does not go to infinity in finite time (just take a smaller $[a, b]$ if this is the case). By definition, $0 = \frac{\partial \rho_t}{\partial t}(z) + \{\rho_t, H\}(z)$. This implies

$$0 = \int_\alpha^\beta \int_A \left( \frac{\partial \rho_t}{\partial t}(z) + \{\rho_t, H\}(z) \right) \xi_t(z) dz dt$$

for any function $\xi_{(\cdot)}(\cdot) : [\alpha, \beta] \times \mathbb{R}^6 \to \mathbb{R}$ almost everywhere continuous, where $[\alpha, \beta] \subset [a, b]$.

Consider then functions $\xi$ vanishing for $t \in \{\alpha, \beta\}$ and for $z \in A$, A being a subset of $\mathbb{R}^6$ not coinciding with $A \cap z[a, b] = \emptyset$. Work in subsets of $[\alpha, \beta]$ and A, use several integration by part to develop this integral to a similar form to the Poisson bracket form of Hamilton's equation. Use the fact that the image of the line in $\mathbb{R}^6$ under any diffeomorphisms is a negligible set in $\mathbb{R}^6$ implies that the same can be done for all subsets of $\mathbb{R}^6$ to deduce that the integrant is zero. Hamilton's equation are then a special case for a trivial $\xi$. Conclude by using a Legendre transform and obtain Newton's equations from the Euler-Lagrange equations. □

## 3.2 The standard Koopman-von Neumann formulation

To develop a hybrid classical-quantum theory, it would be convenient to reformulate the Liouville equation (3.1) for a classical state $\rho \in D := \{\rho : [a, b] \times M \to [0, 1] \,|\, \rho$ *is a smooth and normalized distribution on M smoothly parametrized by a time* $t \in [a, b]\}$ in a framework similar to the one of quantum mechanics, where states are there described as functions in $L^2(M, \mathbb{C})$ parametrized by time. The way that Koopman [24] and von-Neumann [25] achieved the latter was to look for special solutions of the form $\rho = |\chi|^2$ with $\chi \in L^2(R^{2n})$ and study the dynamics of $\chi$.

Interestingly enough, in addition to remarking that these special solutions $\rho$ happen to be all the solution sets, we will show that $\rho$ satisfies Liouville's equation (3.1) if and only if any of its wave function analogue $\chi$ satisfies Liouville's equation (3.1).

Before proceeding, note that we will restrict ourselves to the simple case $\boxed{(M,\omega) = (\mathbb{R}^{2n}, \mathbb{J})}$ for the sake of mathematical lightness. It is not done in this report but the theory could be generalized to arbitrary finite dimensional symplectic connected manifolds by applying the results of the above special case locally on the charts and expanding by using the fact that $M$ is connected. Also, from now on, let us denote $I := [a,b] \subset \mathbb{R}$ where $a \neq b$.

**Proposition 3.2** ("Every $\rho$ admits a classical wave function $\chi$")**.** *For every solution $\rho : [a,b] \times M \to [0,1]$ of the Liouville's equation (3.1), there exists a function $\mathcal{H} \ni \chi : I \times M \to \mathbb{C}$ such that:*

*(i)* $|\chi|^2 = \rho$

*(ii)* $\dfrac{\partial \chi}{\partial t} = \{H, \chi\}$

*where $\rho$ is the probability density function associated with the position of the classical particle.*

*Proof.* Consider $\chi := \exp(i\phi) \cdot \sqrt{\rho}$. Getting $(i)$ is trivial. Part $(ii)$ is a consequence of the linearity of $\{\cdot,\cdot\}$. □

**Proposition 3.3** (Every classical wave function satisfies the Liouville's equation)**.** *On the other side, consider $\mathcal{H} \ni \chi : I \times M \to \mathbb{C}$ such that*

$$\frac{\partial \chi}{\partial t} = \{H, \chi\}. \tag{3.4}$$

*Then, $\rho := |\chi|^2$ satisfies Liouville's equation (3.1)*

*Proof.* One has

$$\begin{aligned} \frac{\partial \rho}{\partial t} &= \frac{\partial \chi}{\partial t} \cdot \chi^* + \chi \cdot \frac{\partial \chi^*}{\partial t} && \text{by definition} \\ &= \{H, \chi\}\chi^* + \chi\{H, \chi^*\} && \text{by (3.4).} \end{aligned} \tag{$\triangle$}$$

On the other side, Jacobi's identity yields

$$\{H, \chi \cdot \chi^*\} = \chi\{H, \chi^*\} + \{H, \chi\}\chi^*. \qquad (\triangle\triangle)$$

Combining ($\triangle\triangle$) in ($\triangle$) yields

$$\begin{aligned} \frac{\partial \rho}{\partial t} &= \{H, \chi\}\chi^* + \chi\{H, \chi^*\} && \text{by } (\triangle) \\ &= \{H, \chi \cdot \chi^*\} - \chi\{H, \chi^*\} + \chi\{H, \chi^*\} && \text{by } (\triangle\triangle) \\ &= \{H, \rho\} && \text{by definition of } \rho \end{aligned}$$

which is what we wanted to prove. □

**Definition 3.3** (Classical wave function)**.** *A function $\chi \in L^2(\mathbb{R}^{2n}, \mathbb{C}; I) := \{\chi : \mathbb{R}^{2n} \times I \to \mathbb{C} \mid \chi(\cdot, t_0) \equiv \chi_{t_0}(\cdot) \in L^2(\mathbb{R}^{2n})\ \forall t_0 \in I,\ \chi(x, \cdot)$ is smooth $\forall x \in \mathbb{R}^{2n}\}$ satisfying the property (i) of proposition (3.2) and a boundary condition at infinity, namely*

$$|\chi|^2 = \rho \qquad \text{(Fundamental prescription)}$$

$$\lim_{|z|\to\infty} \chi_t(z) = 0 \qquad \forall t \in I \qquad \text{(Boundary condition)}$$

*is called a* ***classical wave function****.*

**Remark 3.2** (Set of classical wave functions)**.** *From now on, we will denote the set of classical wave functions by $\mathcal{H}$.*

**Remark 3.3** (State of a system)**.** *One says that a system is in the state $\chi$ if and only if the probability density function related to the position in phase space of the particles in the system is equal to $|\chi|^2$.*

Now that we have shown that Liouville's equation (3.1) admits a Hilbert space formulation for its solutions that are analogue to the quantum theory, we will formulate an operatorial version of Liouville's equation.

### Operatorial expression

We would want to express Liouville equation (3.1) in a similar form as the Schrödinger equation

$$i\hbar\frac{\partial\psi}{\partial t} = \hat{H}\psi, \tag{3.5}$$

*i.e.* we would want to find a classical Hermitian operator $\hat{A} \in \mathrm{Herm}(\mathcal{H})$ such that Liouville's equation (3.1) can be rewritten as $i\kappa\frac{\partial\chi}{\partial t} = \hat{A}\chi$.

In order to do so, start by writing the Liouville's equation for a classical wave function:s formally,

$$\begin{aligned}
i\frac{\partial\chi}{\partial t} = i\{H,\chi\} &= -i\{\chi, H\} && \text{skew symmetry of } \{\cdot,\cdot\} \\
&= -i\nabla\chi\cdot\mathbb{J}\nabla H && \text{canonical Poisson bracket} \\
&= -i\nabla\chi\cdot X_H && \text{by (3.3)} \\
&= -i\kappa\nabla\chi\cdot\kappa^{-1}X_H && \text{for a } \kappa\neq 0 \\
\iff i\kappa\frac{\partial\chi}{\partial t} &= -i\kappa\nabla\chi\cdot X_H. && (3.6)
\end{aligned}$$

Note in passing that the operatorial analogues to the conjugated variables $p$ and $q$ are the classical position and momentum operators defined as follows [14, 26],

**Definition 3.4** (Classical position operator)**.** *The* ***classical position operator*** $\hat{\mathbf{Z}} : \mathcal{H} \to \mathcal{H}^{2n}$ *is defined as* $\hat{\mathbf{Z}}(\chi)[z,t] := z\cdot\chi[x,t]$.

**Definition 3.5** (Classical momentum operator)**.** *The* ***classical momentum operator*** $\hat{\Lambda} : \mathcal{H} \to \mathcal{H}^{2n}$ *is defined as* $\hat{\Lambda}(\chi) := -i\kappa\nabla\chi$.

**Proposition 3.4.** $\hat{\mathbf{Z}}$ *and* $\hat{\Lambda}$ *are conjugated variables, i.e.* $\hat{\mathbf{Z}}$ *and* $\hat{\Lambda}$ *satisfy the following canonical relations*

*(i)* $[\hat{\mathbf{Z}}_m, \hat{\Lambda}_m] = i\kappa\cdot Id \;\; \forall m \in \{1,2,3\}$,

*(ii)* $[\mathbf{Z}_j, \mathbf{Z}_k] = [\hat{\Lambda}_j, \hat{\Lambda}_k] = 0 \;\; \forall j,k \in \{1,2,3\}$.

*Proof.* It follows by direct computation using the definitions. □

Moreover, we achieve our goal by expressing the right hand side of (3.6) as

$$\begin{aligned} i\kappa\frac{\partial\chi}{\partial t} &= -i\kappa\nabla\chi\cdot X_H(\chi) \\ &= X_H(\chi)\cdot\hat{\Lambda}(\chi) \\ &=: \hat{L}_H\chi. \end{aligned} \tag{3.7}$$

**Definition 3.6** (Liouvillian equation of motion). *Equation (3.7) will be refered as the* ***Liouvillian equation of motion****.*

**Definition 3.7** (Koopman-von Neumann transformation and Liouvillian operator). *The* ***Koopman-von Neumann transformation*** $\hat{L} : C^\infty(M,\mathbb{R}) \to \mathrm{Herm}(\mathcal{H})$ *is the map defined by*

$$\hat{L}_H := \hat{L}(H) := X_H\cdot\hat{\Lambda}, \tag{3.8}$$

*where* $\mathrm{Herm}(\mathcal{H})$ *denotes the set of Hermitian operators on* $\mathcal{H}$*. In particular,* $\hat{L}_H$ *is called the* ***Liouvillian operator associated to H****.*

**Remark 3.4** (Interpretation of $\hat{L}_H$). *As* $\hat{H}$ *in (3.5) is the quantum operator describing the total energy of the quantum system and as* $\hat{\Lambda}$ *is meant to be its classical analogue, one would expect* $\hat{L}_H$ *to describe the classical energy.*

**Remark 3.5.** *For dimensional reasons,* $\kappa$ *must have the units of an action. Also,* $\kappa$ *must be independant of the system. Hence, as* $\kappa$ *is the only physical constant satisfying the two above conditions [27], one is naturally led to consider* $\boxed{\kappa = \hbar}$*.*

Now that the basic objects have been defined, one would be interested to express the Liouvillian equation of motion in the Lagrangian and Hamiltonian formalism.

## 3.3 The Hamiltonian setting of the Koopman-von Neumann formulation

It is well known [28] that (3.7) admits a Hamiltonian structure on $\mathcal{H} := L^2(\mathbb{R}^{2n}, \mathbb{C}; I)$ prescribed by the symplectic form

$$\omega(\chi_1,\chi_2) := 2\hbar\,\mathbf{Im}\big[\,\langle\chi_1|\chi_2\rangle\,\big]. \tag{3.9}$$

Equip then $\mathcal{H} := L^2(\mathbb{R}^{2n}, \mathbb{C}; I)$ with its Poisson-bracket

$$\{F, G\} := \frac{1}{2\hbar}\mathbf{Im}\Big[\langle\frac{\delta F}{\delta\chi}|\frac{\delta G}{\delta\chi}\rangle\Big], \tag{3.10}$$

where $F, G : T^*\mathcal{H} \to \mathbb{R}$ as well as with the standard complex $L^2$ inner product defined as

$$\langle\chi_1|\chi_2\rangle := \int_{\mathbb{R}^{2n}} \chi_1^*(z)\chi_2(z)dz \in \mathbb{C}, \tag{3.11}$$

where the real pairing is defined by

$$\langle\chi_1, \chi_2\rangle := \mathbf{Re}\big[\langle\chi_1|\chi_2\rangle\big]. \tag{3.12}$$

**Remark 3.6.** *The link between symplectic form and Poisson-Bracket is made via* (3.2).

Our goals are now (1) to reformulate the Liouvillian equation into a variational principle, and (2) to show that the solutions of the Liouvillian equation (3.7) can be recovered from a variational principle. To do that, we first introduce some new notions of Lagrangian and Hamiltonian before proving the latter.

By remark 3.4, interpret $\hat{L}_H$ as a classical energy operator. Assuming that $\hat{L}_H$ is smooth enough, an inverse Legendre transformation (2.1) of $\hat{L}_H$ yields the following Lagrangian

$$L_{DF}(\chi, \dot{\chi}) := \mathcal{LT}^{-1}[\hat{L}_H](\chi, \dot{\chi}) := \langle\chi, i\hbar\dot{\chi}\rangle - \langle\chi|\hat{L}_H\chi\rangle\,. \tag{3.13}$$

**Definition 3.8** (Dirac-Frankel Lagrangian)**.** *$L_{DF} : \mathcal{H} \to \mathcal{H}$ as defined above is called the **Dirac-Frankel Lagrangian constructed over the operator** $\hat{L}_H$.*

**Definition 3.9.** *The **Hamiltonian functional** $h : \mathcal{H} \to \mathbb{R}$ **associated with an operator** $\hat{H} : \mathcal{H} \to \mathcal{H}$ is defined as*

$$h(\chi) := \langle\chi|\hat{H}\chi\rangle\,. \tag{3.14}$$

**Theorem 3.1** (Variational form of Liouvillian equation)**.** *Assume that the operator $\hat{L}$ satisfies the Cauchy-Lipschitz-Picard-Lindelöf hypotheses, (i.e. is locally bounded and locally uniformly*

*Lipschitz) and is smooth enough such that its inverse Legendre transform* $\mathcal{L}^{-1}[L] : T\mathcal{H}^* \equiv T\mathcal{H} \equiv \mathcal{H} \to T\mathcal{H} \equiv \mathcal{H}$ *exists. Then, solutions* $\chi$ *of the following variational principle of the form* (2.1)

$$\frac{d}{d\epsilon}\Big|_{\epsilon=0} \int_a^b L_{DF}(\chi_\epsilon(t), \dot{\chi}_\epsilon(t))dt = \frac{d}{d\epsilon}\Big|_{\epsilon=0} \int_a^b \langle \chi_\epsilon(t), i\hbar\dot{\chi}\rangle - \langle \chi|\hat{L}\chi\rangle \, dt = 0$$

*are solutions of the Liouvillian equation* (3.7)*, and reciprocally.*

*Proof.* First, remark that the Liouvillian equation (3.7) can be given a similar structure to the one of Hamilton's equations. Indeed,

$$\begin{aligned} i\hbar\frac{\partial\chi}{\partial t} = \hat{L}\chi \iff \frac{\partial\chi}{\partial t} &= -i\hbar^{-1}\hat{L}\chi \\ &= i\hbar^{-1}\mathbb{J}^2\hat{L}\chi \qquad\qquad \text{using } \mathbb{J}^2 = -Id \\ &:= \mathbb{J}\hat{V}(\chi) \end{aligned}$$

where $\hat{V} := i\hbar^1\mathbb{J}\hat{L} : \mathcal{H} \to T\mathcal{H} \equiv \mathcal{H}$ is a vector field on $\mathcal{H}$.

**Lemma 3.1.** $\hat{V}$ *is a Hamiltonian vector field.*

*Proof of the lemma.* $\hat{L}$ satisfies the Cauchy-Lipschitz-Picard-Lindelöf hypothesis, and so does $\hat{V}$. Then, there exists $H : \mathcal{H} \to \mathbb{R}$ such that the solutions of the initial value problem prescribed by $\hat{V}$ are the same that the one prescribed by $X_H$ and reciprocally. Hence, $\hat{V} = X_H$. ■

Conclude by using theorem 2.1. □

So far, the Koopman-von Neumann formalism has been a very natural reformulation of classical mechanics. Unfortunately, we will now see that it carries two fundamental physical limitations. [1].

## Physical limitations

1. **The Dirac-Frankel Lagrangian $L_{DF}$ associated with $\hat{L}_H$ does not transform consistently under local phases**

First, a local phase transformation $T : \mathbb{R}^{2n} \equiv \mathbb{C}^n \to \mathbb{C}^n$ is a transformation of the form $T(z) = \exp(i \cdot F(z))$ where $F : \mathbb{C}^{2n} \to \mathbb{R}$. It is well known [29] that the gauge group for classical mechanics is the group of local phase transformations and that particle Lagrangians are invariant under their group action. However, the Dirac-Frankel Lagrangian (3.13) in the Koopman-von Neumann framework is **not** gauge covariant [27]. As the Koopman-von Neumann formalism is an attempt to reexpress probabilistic mechanics in an operatorial form while capturing all its fundamental results, one would expect gauge-covariance to be one of its features.

2. **The Hamiltonian functional $h$ does not coincide with the total physical energy $H_{tot}$**

   Indeed,

$$
\begin{aligned}
h = \langle \chi | \hat{L}_H \chi \rangle &= \int \chi^*(z) \hat{L}_H \chi(z) dz \\
&= \hbar \int H(z) \cdot \mathbf{Im}[\{\chi^*, \chi\}(z)] dz \\
&\neq \int H(z) \cdot |\chi^2(z)| dz = \int H(z) \cdot \rho(z) dz =: H_{tot},
\end{aligned}
\tag{3.15}
$$

   where the last term is the physical energy of the system.

   The relation (3.15) could inspire one to modify the fundamental prescription $(i)$ of the Koopman-Von Neumann classical wave function stated in proposition 3.2 by setting the condition $\rho = \mathbf{Im}\big[\{\chi, \chi^*\}\big]$ instead of $\rho = |\chi|^2$ [1]. However, this would lead to some inconsistency regarding the probabilistic interpretation of the square of classical wave function as being a normalized density. Indeed,

$$
\int \mathbf{Im}\big[\{\chi^*, \chi\}(z)\big] dz = 0. \tag{3.16}
$$

In his quest to construct a classical-quantum theory admitting a Hamiltonian functional coinciding with the physical energy, Sudarshan [14] had the idea of exploiting the Koopman-Von Neumann formalism and solved the two above physical inconsistencies by invoking special superselection rules. Unfortunately, although extremely valuable, their derivation out of fundamental principles remains unknown and their role unclear [30]. Since then, there have been several attempts [10, 31, 12, 32] to develop a theory like the above-mentioned. This has recently been achieved in the work of Tronci and Gay-Balmaz [1] who considered an alternative formulation that happens to solve the two above problems. Their work will be presented in the following section.

## 3.4 The Koopman-van Hove formulation of classical mechanics

To solve the first problem, Tronci and Gay-Balmaz [1] had the idea of using the minimal coupling method, which is a particular gauge transformation, on the Koopman-von Neumann transformation $\hat{L}$. More precisely, $\hat{L}_H$ has been corrected to

$$\hat{L}_H := X_H \cdot \hat{\Lambda} \longmapsto \hat{\mathcal{L}}_H := \hat{\Phi} + X_H \cdot (\hat{\Lambda} + \hat{\mathcal{A}}), \tag{3.17}$$

where $\hat{\Phi} : \mathcal{H} \to \mathcal{H}$ and $\hat{\mathcal{A}} : \mathcal{H} \to \mathcal{H}^{2n}$ are gauge potential operators.

Physical arguments coming from prequantization theory [33, 34] allow one to fix the gauge potential operators $\Phi$ and $\hat{\mathcal{A}}$ as

$$\Phi := H \cdot \mathbf{Id}, \quad \mathcal{A} := -\frac{1}{2}\mathbb{J}\hat{\mathbf{Z}}. \tag{3.18}$$

Then, (3.18) in (3.17) yields a modified version of the Liouvillian operator. Namely,

$$\begin{aligned}
\hat{\mathcal{L}}_H &\overset{(3.17)}{:=} \Phi + X_H \cdot (\hat{\Lambda} + \hat{\mathcal{A}}) \\
&\overset{(3.18)}{=} H \cdot \mathbf{Id} + X_H \cdot (\hat{\Lambda} - \frac{1}{2}\mathbb{J}\hat{\mathbf{Z}}) \\
&= H \cdot \mathbf{Id} + X_H \cdot -\mathbb{J}(\frac{1}{2}\hat{\mathbf{Z}} - \hat{\Lambda}) \\
&= H \cdot \mathbf{Id} + \mathbb{J}X_H \cdot (\frac{1}{2}\hat{\mathbf{Z}} - \hat{\Lambda}) && \mathbb{J}^2 = -\mathbf{Id} \\
&\overset{(3.3)}{=} H \cdot \mathbf{Id} - \nabla H \cdot (\frac{1}{2}\hat{\mathbf{Z}} - \hat{\Lambda}).
\end{aligned}$$

**Definition 3.10** (Koopman-van Hove transformation and van Hove-Liouvillian operator)**.** *The **Koopman-van Hove transformation** $\hat{\mathcal{L}} : C^\infty(M, \mathbb{R}) \to \mathrm{Herm}(\mathcal{H})$ is the map defined by*

$$\hat{\mathcal{L}}(H) := \hat{\mathcal{L}}_H := H \cdot \mathbf{Id} - \nabla H \cdot \hat{\mathcal{Z}}, \tag{3.19}$$

*where* $\hat{\mathcal{Z}} := \frac{1}{2}\hat{\mathbf{Z}} - \hat{\Lambda} : \mathcal{H} \to \mathcal{H}^{2n}$ *and* $\hat{\mathbf{Z}}$ *is the classical position operator.*
*In particular,* $\hat{\mathcal{L}}_H$ *is called the* ***van Hove-Liouvillian operator****.*

**Remark 3.7** ($\hat{\mathcal{L}}$ is a bijection)**.** *Whereas the transformation* $\hat{L} : C^\infty(M, \mathbb{R}) \to \mathrm{Herm}(\mathcal{H})$ *is surjective but not injective, the transformation* $\hat{\mathcal{L}} : C^\infty(\mathbb{R}, M) \to \mathrm{Herm}(\mathcal{H})$ *is a bijection [1].*

Now that the first physical limitation has been solved with the fundamental correction (3.19), let us investigate the changes that it implies on the rest of the theory, and in particular, on the Liouvillian equation of motion (3.7) and the Hamiltonian functional (3.14).

### 3.4.1 Changes on the Liouvillian equation of motion

**Corollary 3.1** (van Hove Liouvillian equation)**.** *The variation principle (2.1) for the Dirac-Frankel Lagrangian (3.13) associated with the van Hove-Liouvillian operator* $\hat{\mathcal{L}}_H$ *defined in (3.19) yields the following Liouvillian equation*

$$\begin{aligned} i\hbar\frac{\partial \chi}{\partial t} &= \hat{\mathcal{L}}_H \chi \\ &= \{i\hbar H, \chi\} + (H \cdot \mathbf{Id} - \frac{1}{2}\hat{\mathbf{Z}} \cdot \nabla H)\chi. \end{aligned} \tag{3.20}$$

**Remark 3.8** (van Hove-Liouvillian equation)**.** *From now on, the equation (3.20) will be called the* ***van Hove-Liouvillian equation****.*

**Remark 3.9** (Red term)**.** *The red term in the above corollary denotes the extra contribution originating from the van-Hove Liouvillian.*

*Proof.* Develop $\hat{\mathcal{L}}_H$ by using (3.20) and (3.3). Conclude then with theorem (3.1). □

Note that for a special class of Hamiltonian functions, the Koopman-van Hove transformation coincides with the Koopman-von Neumann one.

**Theorem 3.2** (Quadratic Hamiltonians' dynamics remain unchanged)**.** *Consider a quadratic Hamiltonian function* $H \in C^\infty(R^{2n})$*, i.e. a Hamiltonian function of the following form*

$$H(q, p) = \sum_{i=1}^{n} \alpha_i q_i^2 + \beta_i p_i^2. \tag{3.21}$$

*Then,* $\hat{L}_H = \hat{\mathcal{L}}_H$*.*

*Proof.* Observe that $(H \cdot \mathbf{Id} - \frac{1}{2}\hat{\mathbf{Z}} \cdot \nabla H)\chi = (H - H)\chi = 0$. Conclude then by (3.20). □

**Corollary 3.2** (Equations of motion for quadratic Hamiltonian)**.** *Consider a quadratic Hamiltonian $H \in C^\infty(R^{2n})$. Then, the Koopman-von Neumann and Koopman-van Hove equations of motion (3.7) and (3.20) are the same.*

### 3.4.2 Changes on the Hamiltonian functional

**Lemma 3.2** (van Hove-Liouvillian energy)**.** *The hamiltonian functional $h$ associated with the van Hove-Liouvillian operator is*

$$\begin{aligned} h(\chi) &= \int H(z) \cdot \Big( |\chi(z)|^2 + \mathbf{div}(\chi^*(z)\hat{\mathcal{Z}}\chi(z)) \Big) dz \\ &=: \int H(z) \cdot \Big( |\chi(z)|^2 + \mathbf{div}(\hat{\mathcal{J}}\chi) \Big) dz, \end{aligned} \tag{3.22}$$

*where $\hat{\mathcal{J}} : \mathcal{H} \to \mathcal{H}^{2n}$ defined as $\hat{\mathcal{J}}\chi := \chi^*(z)\hat{\mathcal{Z}}\chi(z) \in \mathcal{H}^{2n}$.*

*Proof.* Start with (3.14) and develop by using (3.19). Several integrations by parts lead to the claim. □

Unfortunately, this new Koopman-van Hove transformation (3.19) does not solve the second physical limitation we exposed in the previous section; namely, the Hamiltonian function does still not coincide with the Hamiltonian energy. Indeed, there exists $\chi \in \mathcal{H}$ solution of (3.20) such that $\int H(z)\, \mathbf{div}\big(\hat{\mathcal{J}}\chi\big)(z) dz \neq 0$.

For example, for $n = 1$, take $H(z) := z_1 \cdot z_2$ and $\chi(z,t) := \kappa \exp(\frac{z_2 - z_1}{2} + t)$ on $A := [-1,1]^2$ and 0 outside, where $\kappa \in \mathbb{R}_+$ can be adjusted to satisfy any normalization condition. One can easily convince themself that $\chi$ satisfies the van Hove-Liouvillian equation for $H$ as above. Then, by a direct but lengthy computation

$$\begin{aligned} \int H(z)\, \mathbf{div}\big(\hat{\mathcal{J}}\chi\big)(z) dz &= \int_A z_1 \cdot z_2 \cdot \mathbf{div}\Big( \exp(-z_1 \cdot z_2 \cdot t) \cdot (z_2, z_1) \Big) dz \\ &= \frac{1}{2} \int_A (z_1^2 z_2 + z_1 z_2^2) \cdot \chi(z,t) + \chi(z,t)^2 dz \neq 0. \end{aligned} \tag{3.23}$$

To solve the second physical limitation, Tronci and Gay-Balmaz [1] modified the Koopman-Von Neumann prescription $|\chi|^2 = \rho$ in the definition of a classical wave function, where $\rho : \mathbb{R}^{2n} \to [0,1]$ probability density of the system, into

$$|\chi|^2 + \mathbf{div}(\hat{\mathcal{J}}\chi) = \rho. \tag{3.24}$$

**Proposition 3.5** (Physical consistency of the modified prescription)**.** *The modified prescription* (3.24) *satisfies*

$$h(\chi_{mod}) = \int H(z) \cdot \rho(z) dz, \tag{3.25}$$

*i.e. the Hamiltonian function for the modified wave function coincides with the total physical energy.*

*Proof.* Clear by lemma 3.2. □

The new modifications (3.24) and (3.17) of the standard Koopman-von Neumann theory that solved the two physical limitations lead us to consider a new class of the classical wave functions.

**Definition 3.11** (van Hove classical wave function)**.** *A function* $\chi_{mod} \in L^2(\mathbb{R}^{2n}, \mathbb{C}; I) := \{\chi : \mathbb{R}^{2n} \times I \to \mathbb{C} \mid \chi(\cdot, t_0) \in L^2(\mathbb{R}^{2n}) \ \forall t_0 \in I, \ \chi(x, \cdot)$ *is smooth* $\forall x \in \mathbb{R}^{2n}\}$ *satisfying the modified version of the fundamental prescription* (3.24) *and vanishing at infinity, namely*

$$|\chi_{mod}|^2 + \mathbf{div}(\hat{\mathcal{J}}\chi_{mod}) = \rho \qquad \text{(Modified fundamental prescription)}$$

$$\lim_{|z|\to\infty} \chi_{mod}(z,t) = 0 \qquad \forall t \in I \qquad \text{(Boundary condition)}$$

*where $\rho$ is the probability density function associated with the position of the classical particle at a given position, is called a **van Hove classical wave function**.*

**Remark 3.10** (Set of van Hove wave functions)**.** *From now on, we will denote the set of van Hove classical wave functions by* $\mathcal{H}_{vH}$.

We conclude this chapter by stating a fundamental result concerning the van Hove classical wave function.

**Theorem 3.3** (Liouville equation for $\phi$)**.** *Let* $\phi \in \mathcal{H}_{vH}$. *The quantity* $\rho := |\phi|^2 + \mathbf{div}(\hat{\mathcal{J}}\phi)$ *satisfies Liouville equation.*

*Proof.* Fix a Hamiltonian $H \in C^\infty(\mathbb{R}^{2n})$. Start with $\{H, \rho\}$ and use the fact that $J(\phi) := |\phi|^2 + \mathbf{div}(\hat{\mathcal{J}}\phi)$ is an equivariant momentum map [1]. □

To sum up, the Koopman-van Hove formalism can be summarized in following diagram:

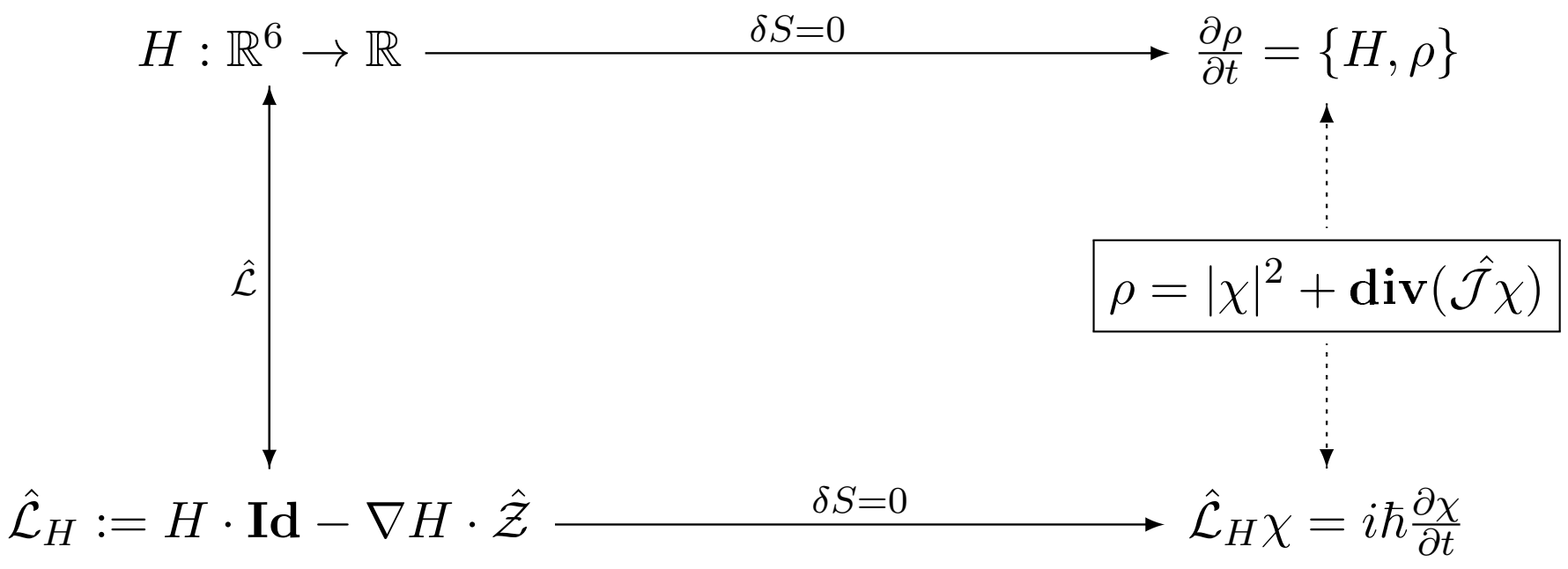

Figure 3.1: Structural diagram of the Koopman-van Hove formulation where $\delta S = 0$ denotes the use of a variational principle. The dashed arrow highlights the relationship that links the two dynamical equations.

# Chapter 4

# Operators in the Koopman-van Hove formulation

So far, we have motivated and exposed the ground theory of the Koopman-van Hove formulation of classical mechanics proposed in [1]. In particular, for a fixed classical probabilistic system, we now know what the analogue of the probability density in the new formalism is (*the van Hove wave function*), how it evolves in time (*according to the van Hove-Liouvillian equation of motion* (3.20)) and how both quantities are related (*through the modified prescription* (3.24)). However, these objects alone are limited as they do not allow, by nature, the description of several fundamental other physical quantities.

In this chapter, we extend the framework by presenting a construction for the operatorial analogues of the classical energy, angular and linear momentum, in this order and in three sections respectively. After each construction, we investigate their respective dynamics and determine the group action generating their averages as momentum maps.

Note that as we have to deal with physics, we will restrict the formalism to $\boxed{n=3}$ for the rest of the report.

## 4.1 The Koopman-von Neumann energy operator

The operatorial analogue to the Hamiltonian in this framework is the van Hove-Liouvillian operator (3.19). Hence, it is natural to consider its associated Hamiltonian functional as the analogue to the total energy of the system. Formally, substituting (3.19) in (3.14) yields the **energy operator** $E : \mathcal{H}_{vH} \to \mathbb{R}$ defined as

$$E(\chi) := h(\chi) := \langle \chi | \hat{\mathcal{L}}_{H_{red}} \chi \rangle \,.$$

In particular, this expression satisfies a conservation law.

**Theorem 4.1** (Energy conservation). *Consider a family of classical wave functions* $\{\chi_t(\cdot)\}_{t\in I}$ $C^2$*- parametrized by time* $t \in I$ *which is solution of the van Hove-Liouvillian equation of motion* (3.20)*. Then* $\frac{\partial}{\partial t}\Big|_{t=t_0} E(\chi_t) = 0$ *for all* $t_0 \in I$*.*

*Proof.* First, remark

**Lemma 4.1.**

$$\hat{\mathcal{L}}_H(\chi_t^*(z))) = -\Big(\hat{\mathcal{L}}_H \chi_t(z))\Big)^* \tag{4.1}$$

*Proof.* Direct computation. ■

Then, by definition,

$$\begin{aligned}
\frac{\partial}{\partial t}\Big|_{t=t_0} E(\chi(z;t)) :&= \frac{\partial}{\partial t}\Big|_{t=t_0} \langle \chi_t | \, \hat{\mathcal{L}}_{H_{red}} \chi_t \rangle \\
&= \int \frac{\partial}{\partial t}\chi_t^*(z) \hat{\mathcal{L}}_H(\chi_t) dz + \int \chi_t^*(z) \hat{\mathcal{L}}_H(\frac{\partial}{\partial t}\chi_t) dz \\
&\overset{(3.20),(4.1)}{=} -\frac{1}{i\hbar}\int \Big(\hat{\mathcal{L}}_H(\chi_t)\Big)^* \hat{\mathcal{L}}_H(\chi_t) dz + \frac{1}{i\hbar}\int \chi_t^*(z)\Big(\hat{\mathcal{L}}_H \circ \hat{\mathcal{L}}_H\Big)(\chi_t) dz \\
&\overset{(3.11)}{=} \frac{1}{i\hbar}\Big( - \langle \hat{\mathcal{L}}_H(\chi_t) | \, \hat{\mathcal{L}}_H(\chi_t) \rangle + \langle \chi_t | \, (\hat{\mathcal{L}}_H \circ \hat{\mathcal{L}}_H)(\chi_t) \rangle \Big) \\
&= \frac{1}{i\hbar}\Big( - \langle \hat{\mathcal{L}}_H(\chi_t) | \, \hat{\mathcal{L}}_H(\chi_t) \rangle + \langle \hat{\mathcal{L}}_H(\chi_t) | \, \hat{\mathcal{L}}_H(\chi_t) \rangle \Big) \\
&= 0.
\end{aligned}$$

□

Now that we dealt with the notion of energy in the Koopman-van Hove formalism, we would want to construct operatorial analogues $\hat{\mathbb{L}}, \hat{\mathbb{P}} : \hat{\mathcal{H}} \to \mathcal{H}^3$ to the angular and linear momentum and study their properties from a geometrical perspective. This will be done, in this order, in the next 2 sections.

## 4.2 The Koopman-van Hove angular momentum operator

First, as the classical angular momentum $L$ is defined as $L := p \times q$ and as we have seen that the conjugated operators $\hat{Z}$ and $\hat{\Lambda}$ are the analogues to the conjugated variables $q$ and $p$, it would be natural to investigate a definition of the form

$$\hat{\mathbb{L}}' = \hat{Z} \times \hat{\Lambda} \text{ where } \hat{\mathbb{L}}'\chi(z) = \chi(z) \begin{pmatrix} q \\ p \end{pmatrix} \times (-i\hbar) \begin{pmatrix} \nabla_q \chi(z) \\ \nabla_p \chi(z) \end{pmatrix}.$$

Although natural, some issues with the above proposition need to be solved. First, one has to give sense to the notion of cross-product between vectors in $\mathbb{R}^6$. Second, to be analogue to the classical angular momentum, one would want need to transform $\hat{\mathbb{L}}' : \mathcal{H} \to \mathcal{H}^6$ to a $\hat{\mathbb{L}}'' : \mathcal{H} \to \mathcal{H}^3$. Although these 2 issues could be fixed by a smart use of matrices and commutators and by using quantum operator algebra [27], we will propose a more fundamental construction.

### 4.2.1 Construction of $\hat{\mathbb{L}}$

The fundamental idea of this construction will use the fact that the map $C^\infty(\mathbb{R}) \ni H \longmapsto \hat{\mathcal{L}}_H \in \text{Herm}(\mathcal{H})$ between the Hamiltonian functions and the set of Hermitian operators is a Lie algebra isomorphism [1, 27]. As we want our operator $\hat{\mathbb{L}}$ to be Hermitian to fit in our framework, by the above result, there must exists a Hamiltonian function $H$ such that $\hat{\mathcal{L}}_H = \hat{\mathbb{L}}_\xi$. The question is now to find the right Hamiltonian $H$. To do that, we will expose the Lie algebra structure emerging in the geometrical mechanical derivation of the angular momentum [16] and outline a fundamental Hamiltonian quantity.

Consider the action of $SO(3)$ on $T^*\mathbb{R}^3 \simeq \mathbb{R}^6$ given by $SO(3) \times T^*\mathbb{R}^3 \ni (A, (q,p)) \mapsto (Aq, Aq)$, consisting in a rotation of coordinates. We are interested in determining the momentum map associated with this action. Let $\tilde{\xi} \in \mathfrak{so}(3)$ be an initial velocity in the Lie algebra $\mathfrak{so}(3) \equiv T_e SO(3)$. Then, as this action is a matrix Lie group action [16], the infinitesimal generator $\hat{\xi}_{\mathbb{R}^6}$ is simply $\hat{\xi}_{\mathbb{R}^6}(q,p) = (\tilde{\xi}q, \hat{\xi}p) = (\xi \times q,\, \xi \times p)$, where we used the hat-map isomorphism between $\mathfrak{so}(3) = \{A \in SO(3) \mid A^T + A = 0\}$ and $\mathbb{R}^3$.

Finally, by definition 3.3 in the special case of $(M, \omega) = (\mathbb{R}^6, \mathbb{J})$, the restricted momentum map $J_{\hat{\xi}}$ associated with the above action satisfies the following PDE.:

$$X_{J_{\hat{\xi}}}(q,p) = \hat{\xi}_{\mathbb{R}^6}(q,p) \overset{(3.3)}{\Longleftrightarrow} \left(\frac{\partial J_{\hat{\xi}}}{\partial p}(q,p), -\frac{\partial J_{\hat{\xi}}}{\partial q}(q,p)\right) = (\xi \times q,\, \xi \times p),$$

whose solution $J_{\hat{\xi}}$ is

$$J_{\hat{\xi}}(q,p) = (q \times p) \cdot \xi. \tag{4.2}$$

Finally, as the momentum map $J$ is a map from : $\mathbb{R}^6 \to \mathfrak{so}(3)^*$, $J(p,q) \in \mathfrak{so}(3)^*$, the identification $\mathfrak{so}(3)^* \ni J(q,p) \equiv \mathbf{J}(q,p) \in \mathbb{R}^3$ prescribed by the breve map [16] allows us to write

$$J_{\hat{\xi}}(q,p) = \langle \mathbf{J}(q,p), \xi \rangle \overset{(4.2)}{=} (q \times p) \cdot \xi.$$

By non-degeneracy of the pairing, we conclude that $J(q,p) := q \times p$. In other words, the momentum map associated with the action of rotations on $\mathbb{R}^6$ is the classical angular momentum.

Note that $J_{\hat{\xi}}$ is fundamentally, and by construction, a necessary object to determine the momentum map $J(q,p) = q \times p =: L$. Moreover, the restricted momentum map $J_{\hat{\xi}} : \mathbb{R}^6 \to \mathbb{R}$ is a Hamiltonian function. Motivated by the above two facts, we will consider $H = J_\xi$ and construct an operator $\hat{\mathbb{L}}_\xi := \hat{\mathcal{L}}_H$ describing the projection of the angular momentum operator on a vector $\xi$ before extending it to a general $\hat{\mathbb{L}}$.s

### 4.2.2 Computation of the angular momentum operator $\hat{\mathbb{L}}_\xi$ and extension

As motivated above, consider the Hamiltonian $J_\xi(q,p) := q \times p \cdot \xi$. By definition of the van Hove-Liouvillian operator (3.19), one has

$$\begin{aligned}
\hat{\mathbb{L}}_\xi := \hat{\mathcal{L}}_{J_\xi} &\overset{(3.19)}{=} \hbar\{iJ_\xi, \cdot\} - \left(\begin{pmatrix} -\xi \times p \\ \xi \times q \end{pmatrix} \cdot \frac{1}{2}\hat{\mathbf{Z}} - J_\xi\right) \\
&= \hbar\{iJ_\xi, \cdot\} - \left( -\frac{1}{2}(\xi \times p) \cdot \hat{\mathbf{Q}} + \frac{1}{2}(\xi \times q) \cdot \hat{\mathbf{P}} - \xi \cdot q \times p \cdot \mathbf{Id}\right) \\
\Longleftrightarrow \hat{\mathbb{L}}_\xi(\chi)(z) &= \hbar\{iJ_\xi, \chi\}(z) + \cancel{\frac{1}{2}(\xi \times p) \cdot q} \cdot \chi(z) - \cancel{\frac{1}{2}(\xi \times q) \cdot p} \cdot \chi(z) + \cancel{\xi \cdot q \times p \cdot \chi(z)}
\end{aligned}$$

$$
\begin{aligned}
&= i\hbar\Big((p \times \xi)\cdot \frac{\partial}{\partial p}\chi(z) - (\xi \times q)\cdot \frac{\partial}{\partial q}\chi(z)\Big)\\
\iff \hat{\mathbb{L}}_\xi(\chi)(z) &= i\hbar\Big(\frac{\partial}{\partial p}\chi(z) \times p + \frac{\partial}{\partial q}\chi(z)\times q\Big)\cdot \xi, \qquad (4.3)
\end{aligned}
$$

where we used the property $(a \times b)\cdot c = (b \times c)\cdot a$.

**Remark 4.1.** *Note that for the angular momentum projection $J_\xi$, $\hat{\mathcal{L}}_{J_\xi} = \hat{L}_{J_\xi}$, i.e. this construction for the Koopman-von Neumann and Koopman-van Hove transformations yield the same operator on $J_\xi$.*

As $\hat{\mathbb{L}}_\xi : \mathcal{H} \to \mathcal{H}$ translates the projection of the angular momentum on a vector $\xi \in \mathbb{R}^3$, one is naturally led to define the following extension.

**Definition 4.1** (Koopman-van Hove angular momentum operator)**.** *The Koopman-van Hove angular momentum operator $\hat{\mathbb{L}}$ is the operator $\hat{\mathbb{L}} : \mathcal{H}_{vH} \to \mathcal{H}^3_{vH}$ defined as*

$$
\hat{\mathbb{L}}(\chi)(z) = i\hbar\Big(\frac{\partial}{\partial p}\chi(z) \times p + \frac{\partial}{\partial q}\chi(z)\times q\Big). \qquad (4.4)
$$

### 4.2.3 Dynamics of $\hat{\mathbb{L}}$ and $\langle\hat{\mathbb{L}}\rangle$

The dynamics of the Koopman-van Hove angular momentum operator $\hat{\mathbb{L}}$ as well as its average $\langle\hat{\mathbb{L}}\rangle$ are prescribed as follows.

**Theorem 4.2** (Dynamics of $\hat{\mathbb{L}}$ and $\langle\hat{\mathbb{L}}\rangle$)**.** *Let $\hat{\mathbb{L}}$ be the Koopman-van Hove angular momentum operator. Consider a solution $\chi_t \in \mathcal{H}_{vH}$ of the van Hove-Liouvillian equation (3.20). Then*

*(i)* $\displaystyle \frac{d}{dt}\hat{\mathbb{L}}(\chi_t)(z) = \Big(\frac{\partial}{\partial q}\hat{\mathcal{L}}_H(\chi_t)(z)\times q + \frac{\partial}{\partial p}\hat{\mathcal{L}}_H(\chi_t)(z)\times p\Big) = -\frac{i}{\hbar}\Big(\hat{\mathbb{L}}\circ\hat{\mathcal{L}}_H\Big)(\chi_t)(z),$

*(ii)* $\displaystyle \frac{d}{dt}\langle\hat{\mathbb{L}}\rangle(\chi_t) = \frac{2}{\hbar}\,\mathbf{Im}\Big[\langle\hat{\mathbb{L}}_\xi(\chi_t)|\,\hat{\mathcal{L}}_H(\chi_t)\rangle\Big].$

*Proof.* For (i), use smoothness of $\chi_t$ and apply definition (3.7).

For (ii), pick a $\xi \in \mathbb{R}^3$. By definition

$$
\langle\hat{\mathbb{L}}_\xi\rangle(\chi) := \langle\chi\,|\,\hat{\mathbb{L}}_\xi\chi\rangle
$$

$$= i\hbar \int \chi^*(z) \Big( \frac{\partial}{\partial p} \chi(z) \times p + \frac{\partial}{\partial q} \chi(z) \times q \Big) \cdot \xi dz. \tag{4.5}$$

Then, using lemma (4.1),

$$\begin{aligned}
\frac{d}{dt} \langle \hat{\mathbb{L}}_\xi \rangle (\chi_t) &= \int \frac{\partial}{\partial t} \chi_t^*(z) \hat{\mathbb{L}}_\xi(\chi_t(z)) + i\hbar \chi_t^*(z) \Big( \frac{\partial}{\partial t} \frac{\partial}{\partial q} \chi_t(z) \times q + \frac{\partial}{\partial t} \frac{\partial}{\partial p} \chi_t(z) \times p \Big) \cdot \xi dz \\
&\overset{(4.1)}{=} \int -\frac{1}{i\hbar} \Big( \hat{\mathcal{L}}_H(\chi_t)(z) \Big)^* \hat{\mathbb{L}}_\xi(\chi_t(z)) dz + \frac{1}{i\hbar} \chi_t^*(z) \hat{\mathbb{L}}_\xi \Big( \hat{\mathcal{L}}_H(\chi)(z) \Big) \cdot \xi dz \\
&\overset{(3.12)}{=} -\frac{1}{i\hbar} \langle \hat{\mathcal{L}}_H(\chi_t) | \hat{\mathbb{L}}_\xi(\chi_t) \rangle + \frac{1}{i\hbar} \langle \chi_t | \hat{\mathbb{L}}_\xi \Big( \hat{\mathcal{L}}_H(\chi_t) \Big) \rangle \\
&= -\frac{1}{i\hbar} \langle \hat{\mathcal{L}}_H(\chi_t) | \hat{\mathbb{L}}_\xi(\chi_t) \rangle + \frac{1}{i\hbar} \langle \hat{\mathbb{L}}_\xi(\chi_t) | \hat{\mathcal{L}}_H(\chi_t) \rangle \qquad \text{as } \hat{\mathbb{L}}_\xi \in \mathrm{Herm}(\mathcal{H}_{vH}) \\
&= \frac{1}{i\hbar} \Big( - \langle \hat{\mathcal{L}}_H(\chi_t) | \hat{\mathbb{L}}_\xi(\chi_t) \rangle + \langle \hat{\mathbb{L}}_\xi(\chi_t) | \hat{\mathcal{L}}_H(\chi_t) \rangle \Big) \qquad \text{use skew-symmetry of the } L^2 \text{ pairing} \\
&= \frac{2}{\hbar} \mathbf{Im} \Big[ \langle \hat{\mathbb{L}}_\xi(\chi_t) | \, \hat{\mathcal{L}}_H(\chi_t) \rangle \Big].
\end{aligned}$$

As this is true for any $\xi \in \mathbb{R}^3$, we have that $\frac{d}{dt} \langle \hat{\mathbb{L}} \rangle = \frac{2}{\hbar} \mathbf{Im} \Big[ \langle \hat{\mathbb{L}}(\chi_t), \, \hat{\mathcal{L}}_H(\chi_t) \rangle \Big]$. □

**Corollary 4.1** (Extra contributions)**.** *Fix a Hamiltonian $H \in C^\infty(\mathbb{R}^6)$. For $\hat{\mathcal{L}}_H = \hat{L}_H + \alpha_H$ where $\alpha_H := -\frac{1}{2} \hat{\mathbf{Z}} \cdot \nabla H + H$, we have*

*(i)* $\frac{d}{dt} \hat{\mathbb{L}}(\chi_t)(z) = -\frac{i}{\hbar} \Big( \hat{\mathbb{L}} \circ \hat{L}_H \Big)(\chi_t)(z) - \frac{i}{\hbar} \Big( \hat{\mathbb{L}} \circ \alpha_H \Big)(\chi_t)(z),$

*(ii)* $\frac{d}{dt} \langle \hat{\mathbb{L}} \rangle (\chi_t) = \frac{2}{\hbar} \mathbf{Im} \Big[ \langle \hat{\mathbb{L}}(\chi_t) | \, \hat{L}_H(\chi_t) \rangle \Big] + \frac{2}{\hbar} \mathbf{Im} \Big[ \langle \hat{\mathbb{L}}(\chi_t) | \, \alpha_H(\chi_t) \rangle \Big].$

**Remark 4.2.** *In the standard formalism, the dynamics would have consisted only in the first term. The red term are the new contributions appearing in the modified framework.*

### 4.2.4 Underlying Lie group action

We now want to get a deeper structural understanding of the conservation law associated with the average of the Koopman-van Hove angular momentum operator $\langle \hat{\mathbb{L}} \rangle$ from a geometrical mechanical perspective. Similarly to the description of the classical angular momentum $L = q \times p$ done in section 4.2.1, we are interested to determine a group action on $\mathcal{H}_{vH}$ such that its associated momentum map $J$ is $\langle \hat{\mathbb{L}} \rangle$.

First, assuming that what we are looking for exists, seeing $\langle \hat{\mathbb{L}} \rangle : \mathbb{R}^6 \to \mathbb{R}^3$ as a momentum map implies, by definition, that the Lie algebra $\mathfrak{g}$ of the Lie group $G$ that we are looking for satisfies $\mathbb{R}^3 \simeq \mathfrak{g}^* \simeq \mathfrak{g}$. Consider then $\xi \in \mathbb{R}^3$ and a $\chi \in \mathcal{H}_{vH}$. By proposition 2.1, we have

$$\begin{aligned} \langle \hat{\mathbb{L}} \rangle \text{ is a momentum map : } &\overset{Def.\ 2.6}{\Longleftrightarrow} X_{\langle \hat{\mathbb{L}} \rangle}(\chi) = \xi_{\mathcal{H}_{vH}}(\chi) \\ &\overset{Prop.\ 2.1}{\Longleftrightarrow} \langle \langle \hat{\mathbb{L}} \rangle (\chi), \xi \rangle_{\mathfrak{g}^* \times \mathfrak{g}} = \hbar \langle i\xi_{\mathcal{H}_{vH}}(\chi), \chi \rangle_{T\mathcal{H} \times \mathcal{H}}, \end{aligned} \tag{4.6}$$

where $\langle \cdot, \cdot \rangle_{\mathfrak{g}^* \times \mathfrak{g}}$ and $\langle \cdot, \cdot \rangle_{T\mathcal{H} \times \mathcal{H}} \equiv \langle \cdot, \cdot \rangle_{\mathcal{H}} \equiv \langle \cdot, \cdot \rangle$ are respectively the natural pairing on the Lie algebra and the $L^2$ inner product defined in (3.12). More specifically, in our case, the bilinear operator $\langle \cdot, \cdot \rangle_{\mathfrak{g}^* \times \mathfrak{g}}$ is the standard Euclidean inner product. Then, equation (4.6) becomes

$$\langle \langle \hat{\mathbb{L}} \rangle (\chi), \xi \rangle_{\mathfrak{g}^* \times \mathfrak{g}} = \hbar \langle i\xi_{\mathcal{H}_{vH}}(\chi), \chi \rangle_{T\mathcal{H} \times \mathcal{H}} \iff \langle \hat{\mathbb{L}}_\xi \rangle (\chi) = \hbar \langle i\xi_{\mathcal{H}_{vH}}(\chi), \chi \rangle \in \mathbb{R}. \tag{4.7}$$

Let us work $\langle \hat{\mathbb{L}}_\xi \rangle (\chi)$ before coming back in (4.7).

$$\begin{aligned} \langle \hat{\mathbb{L}}_\xi \rangle (\chi) &\overset{(4.5)}{:=} \hbar \int \chi^*(z) i \Big[ \xi \cdot \Big( \frac{\partial}{\partial q}(\cdot) \times q + \frac{\partial}{\partial p}(\cdot) \times p \Big) \Big] \chi(z) dz \\ &= \hbar \int \chi^*(z)\, i\hat{\mathbf{A}}_\xi \chi(z) dz \\ &= \hbar \langle \chi | i\hat{\mathbf{A}}_\xi \chi \rangle \in \mathbb{C} \qquad \text{where } \hat{\mathbf{A}}_\xi := \xi \cdot \Big( \frac{\partial}{\partial q}(\cdot) \times q + \frac{\partial}{\partial p}(\cdot) \times p \Big). \end{aligned} \tag{4.8}$$

In particular, as (4.7) and (4.8) are equal and (4.7) is real, we have that

$$\hbar \langle \chi | i\hat{\mathbf{A}}_\xi \chi \rangle = \hbar \langle \chi, i\hat{\mathbf{A}}_\xi \chi \rangle \in \mathbb{R}. \tag{4.9}$$

Hence, equation (4.7) becomes

$$\begin{aligned} &\hbar \langle i\xi_{\mathcal{H}_{vH}}(\chi), \chi \rangle \overset{(4.8),(4.9)}{=} \hbar \langle \chi, i\hat{\mathbf{A}}_\xi \chi \rangle \\ &\iff \langle i\xi_{\mathcal{H}_{vH}}(\chi), \chi \rangle = \langle i\hat{\mathbf{A}}_\xi \chi, \chi \rangle \\ &\iff \langle i(\xi_{\mathcal{H}_{vH}} - \hat{\mathbf{A}}_\xi)\chi, \chi \rangle = 0 && \forall \chi \in \mathcal{H}_{vH} \\ &\Rightarrow \xi_{\mathcal{H}_{vH}} - \hat{\mathbf{A}}_\xi = 0 && \text{By non-degeneracy of the pairing} \\ &\iff \hat{\mathbf{A}}_\xi = \xi_{\mathcal{H}_{vH}} \overset{(2.11)}{:=} \frac{d}{dt}\Big|_{t=0} \exp(t\xi). \end{aligned} \tag{4.10}$$

The question of finding a group action on $\mathcal{H}_{vH}$ such that $\langle\hat{\mathbb{L}}\rangle$ is a momentum map is now equivalent to finding a group action satisfying (4.10) for all $\xi \in \mathbb{R}^3$.

**Theorem 4.3** (Group action generating $\langle\hat{\mathbb{L}}\rangle$ as a momentum map)**.** *Consider $G = SO(3)$ and the action $\Phi : SO(3) \times \mathcal{H}_{vH} \to \mathcal{H}_{vH}$ on $\mathcal{H}_{vH}$ defined by*

$$\Phi(A, \chi)(z) := \chi(-Aq, -Ap) \in \mathbb{C}. \tag{4.11}$$

*Then, $\langle\hat{\mathbb{L}}\rangle$ is the momentum map associated with the action $\Phi$ of $SO(3)$ on $\mathcal{H}_{vH}$.*

*Proof.* Fix $\xi \in \mathfrak{g} \equiv \mathbb{R}^3$. Let $\{g_t\}_{t\in I} \subset SO(3)$ be a 1-parameter family of group elements in $SO(3)$ such that $g(0) = \mathbf{Id}$ and $g'(0) = \xi$.

Then,

$$\begin{aligned}
\xi_{\mathcal{H}_{vH}}(\chi)(z) &\overset{(2.11)}{=} \left.\frac{d}{dt}\right|_{t=0} \exp(t\xi)\chi(z) \\
&\overset{(4.11)}{=} \left.\frac{d}{dt}\right|_{t=0} \Big(\chi(-g(t)q, -g(t)p)\Big) \\
&= \frac{\partial\chi}{\partial q} \cdot (-\xi \times q) + \frac{\partial\chi}{\partial p} \cdot (-\xi \times p) \\
&= \xi \cdot \left(\frac{\partial\chi}{\partial q} \times q + \frac{\partial\chi}{\partial p} \times p\right) \\
&\overset{!}{=} \hat{\mathbf{A}}_\xi(\chi)(z)
\end{aligned}$$

$\overset{(4.10)}{\iff}$ $\Phi$ is indeed the group action generating $\langle\hat{\mathbb{L}}\rangle$ as a momentum map. □

## 4.3 The Koopman-van Hove linear momentum operator

In this section, we will construct the Koopman-van Hove analogue to the linear momentum, investigate its dynamics and determine the group action generating its average as a momentum map.

### 4.3.1 Construction of $\hat{\mathbb{P}}$

The Koopman-van Hove linear momentum operator $\hat{\mathbb{P}}$ will be constructed in a similar fashion to $\hat{\mathbb{L}}$; namely by using the Lie algebra isomorphism $\hat{\mathcal{L}}$ for a good choice of $H$. To do that, we will expose the Lie algebra structure emerging when describing the classical linear momentum as a momentum map.

Consider the action of $\mathbb{R}^3$ on $T^*\mathbb{R}^3 \simeq \mathbb{R}^6$ given by $\mathbb{R}^3 \times T^*\mathbb{R}^3 \ni (v,(q,p)) \mapsto (q+v,p) \in T^*\mathbb{R}^3$, basically describing a translation of the positions of the bodies. We are interested in determining the momentum map associated with this action. Let $\xi \in \mathbb{R}^3$ be an initial velocity in the Lie algebra of $\mathbb{R}^3$. Consider now the family $\{v_t\}_{t\in[-a;a]} \subset \mathbb{R}^3$, $a \in \mathbb{R}\setminus\{0\}$, such that $v_0 = Id$ and such that $\frac{d}{dt}\Big|_{t=0} v_t = \xi$. By definition, the infinitesimal generator $\xi_{\mathbb{R}^3}$ is

$$\xi_{\mathbb{R}^6}(q,p) \overset{(2.11)}{=} \frac{d}{dt}\Big|_{t=0}\Big(\exp(t\xi)z\Big) = \frac{d}{dt}\Big|_{t=0}(q+v_t,p) = (\xi,0).$$

Finally, by definition 2.6 in the special case of $(M,\omega) = (\mathbb{R}^6,\mathbb{J})$, the restricted momentum map $J_{\hat{\xi}}$ associated with the above action satisfies the following PDE

$$X_{J_{\hat{\xi}}}(q,p) = \xi_{\mathbb{R}^6}(q,p) \overset{(3.3)}{\iff} \left(\frac{\partial J_{\hat{\xi}}}{\partial p}(q,p), -\frac{\partial J_{\xi}}{\partial q}(q,p)\right) = (\xi,\ 0),$$

whose solution $J_{\hat{\xi}}$ is $J_\xi(q,p) = p\cdot\xi$. As this is true for arbitrary $\xi$, $J_\xi$ can be extended to

$$J(q,p) = p. \tag{4.12}$$

### 4.3.2 Computation of the linear momentum operator $\hat{\mathbb{P}}_\xi$ and extension

For $H = J_\xi$ as above and by definition of the van Hove-Liouvillian operator, one finds

$$\begin{aligned}\hat{\mathbb{P}}_\xi := \hat{\mathcal{L}}_{J_\xi} &\overset{(3.19)}{=} \{i\hbar J_\xi,\cdot\} - \left(\frac{1}{2}\hat{\mathbf{Z}}\cdot\nabla J_\xi - J_\xi\right)\\ &= i\hbar\left(\frac{\partial J_\xi}{\partial q}\frac{\partial}{\partial p} - \frac{\partial J_\xi}{\partial p}\frac{\partial}{\partial q}\right) - \left(\frac{1}{2}\hat{\mathbf{P}}\cdot\xi - \hat{\mathbf{P}}\cdot\xi\right)\mathbf{Id}\end{aligned}$$

$$= -i\hbar\xi \cdot \frac{\partial}{\partial q} + \frac{1}{2}\hat{\mathbf{P}} \cdot \xi\, \mathbf{Id} \tag{4.13}$$

$$\iff \hat{\mathbb{P}}_\xi \chi(z) = \Big( -i\hbar\frac{\partial \chi}{\partial q}(z) + \frac{1}{2} p\, \chi(z) \Big) \cdot \xi. \tag{4.14}$$

Moreover, the operator $\hat{\mathbb{P}}_\xi$ being understood as the projection of the linear operator on the vector $\xi$, one is naturally led to define $\hat{\mathbb{P}}$ as follows.

**Definition 4.2** (Koopman-van Hove linear momentum operator)**.** *The Koopman-van Hove linear momentum operator $\hat{\mathbb{P}}$ is the operator $\hat{\mathbb{P}} : \mathcal{H} \to \mathcal{H}^3$ defined as*

$$\hat{\mathbb{P}}\chi(z) = -i\hbar\frac{\partial \chi}{\partial q}(z) + \frac{1}{2} p\, \chi(z). \tag{4.15}$$

**Remark 4.3.** *Note that in this case, $\hat{L}_{J_\xi} := -i\hbar\frac{\partial \chi}{\partial q}(z) \neq -i\hbar\frac{\partial \chi}{\partial q}(z) + \frac{1}{2}p\,\chi(z) =: \hat{\mathcal{L}}_{J_\xi}\chi(z)$. In words, the same constructions by using the Koopman-von Neumann and Koopman-van Hove transformations yield different operators for $J_\xi$.*

Similarly, and to understand the new contributions appearing in the modified theory, we will need to define the standard equivalent of $\hat{\mathbb{P}}$.

**Definition 4.3** (Standard Koopman-von Neumann linear momentum operator)**.** *We define the* ***standard Koopman-von Neumann linear momentum operator*** *$\hat{\mathbb{P}}^{KvN} : \mathcal{H}_{vH} \to \mathcal{H}_{vH}$ the same way we did for the modified formalism but using instead the standard Koopman-von Neumann transformation (3.8). Formally,*

$$\hat{\mathbb{P}}^{KvN} := \hat{L}_{J_\xi} \qquad \textit{for } J_\xi(q,p) := p \cdot \xi. \tag{4.16}$$

Now that we obtained an explicit expression for the linear momentum operator in the Koopman-van Hove formalism, we are interested to investigate the dynamics of $\hat{\mathbb{P}}$ and $\langle\hat{\mathbb{P}}\rangle$, and more importantly, if they are conserved quantities.

### 4.3.3 Dynamics of $\hat{\mathbb{P}}$ and $\langle\hat{\mathbb{P}}\rangle$

Recall that the deterministic classical case, the total linear momentum is a conserved quantity if the system does not exchange matter with the surroundings **and** is not acted by external forces.

In our case, the framework is such that only the first condition is satisfied as the dimension of the manifold is fixed. Concerning the second condition, our choice of Hamiltonian $H \in C^\infty(M;\mathbb{R})$ is such that it does encompass systems whose dynamics includes external forces. Hence, to be a physically consistent theory, one would expect $\hat{\mathbb{P}}$ and $\langle\hat{\mathbb{P}}\rangle$ to **not** be conserved quantities in general. As expected, we will see that they are not.

**Theorem 4.4** (Dynamics of $\hat{\mathbb{P}}$ and $\langle\hat{\mathbb{P}}\rangle$)**.** *Consider a solution $\chi_t \equiv \chi(t,\cdot) \in \mathcal{H}_{vH}$ of the van Hove-Liouvillian equation* (3.20) *associated with a Hamiltonian $H$. Then:*

*(i)* $\dfrac{d}{dt}\hat{\mathbb{P}}(\chi_t)(z) = -\Big(\dfrac{\partial}{\partial q} + \dfrac{i}{2\hbar}p\Big)\hat{\mathcal{L}}_H(\chi_t)(z) = -\dfrac{i}{\hbar}\Big(\hat{\mathbb{P}}\circ\hat{\mathcal{L}}_H\Big)(\chi_t)(z),$

*(ii)* $\dfrac{d}{dt}\left\langle\hat{\mathbb{P}}\right\rangle(\chi_t)dt = \dfrac{2}{\hbar}\mathbf{Im}\Big[\langle\,\hat{\mathbb{P}}(\chi_t)|\hat{\mathcal{L}}_H(\chi_t)\rangle\Big].$

*Proof.* For (i), start with the definition and use the smoothness of $\chi_t$. Concerning (ii), recall that $\hat{\mathbb{P}}$ is Hermitian by construction. Then, by definition,

$$\begin{aligned}
\frac{d}{dt}\left\langle\hat{\mathbb{P}}\right\rangle(\chi_t)dt &= -i\hbar\int\frac{\partial}{\partial t}\chi_t^*(z)\Big(\frac{\partial}{\partial q}+\frac{i}{2\hbar}p\Big)\chi_t(z)dz - i\hbar\int\chi_t^*(z)\Big(\frac{\partial}{\partial q}+\frac{i}{2\hbar}p\Big)\frac{\partial}{\partial t}\chi_t(z)dz\\
&\overset{(3.20),(4.1)}{=}\int\Big(\hat{\mathcal{L}}_H(\chi_t)(z)\Big)^*\Big(\frac{\partial}{\partial q}+\frac{i}{2\hbar}p\Big)\chi_t(z)dz + \frac{1}{i\hbar}\int\chi_t^*(z)\hat{\mathbb{P}}\Big(\hat{\mathcal{L}}_H(\chi_t)(z)\Big)dz\\
&= \frac{i}{\hbar}\langle\hat{\mathcal{L}}_H(\chi_t)|\,\hat{\mathbb{P}}(\chi_t)\rangle - \frac{i}{\hbar}\langle\chi_t|\,\hat{\mathbb{P}}(\hat{\mathcal{L}}_H(\chi_t))\rangle\\
&= \frac{2}{\hbar}\mathbf{Im}\Big[\langle\,\hat{\mathbb{P}}(\chi_t)|\hat{\mathcal{L}}_H(\chi_t)\rangle\Big].
\end{aligned}$$

□

**Corollary 4.2** (Extra contributions)**.** *Fix a Hamiltonian $H \in C^\infty$. For $\hat{\mathcal{L}}_H = \hat{L}_H + \alpha_H$ where $\alpha_H := -\frac{1}{2}\mathbf{Z}\cdot\nabla H + H$, we have*

*(i)*
$$\begin{aligned}
\frac{d}{dt}\hat{\mathbb{P}}(\chi_t)(z) = &-\frac{i}{\hbar}\Big(\hat{\mathbb{P}}^{KvN}\circ\hat{L}_H\Big)(\chi_t)(z) - \frac{i}{\hbar}\Big(\hat{\mathbb{P}}^{KvN}\circ\alpha_H\Big)(\chi_t)(z)\\
&-\frac{i}{2\hbar}\Big(\hat{\mathbf{P}}\circ\hat{L}_H\Big)(\chi_t)(z) - \frac{i}{2\hbar}\Big(\hat{\mathbf{P}}\circ\alpha_H\Big)(\chi_t)(z),
\end{aligned}$$

*(ii)*
$$\begin{aligned}
\frac{d}{dt}\left\langle\hat{\mathbb{P}}\right\rangle(\chi_t)dt = &\frac{2}{\hbar}\mathbf{Im}\Big[\langle\,\hat{\mathbb{P}}^{KvN}(\chi_t)|\hat{L}_H(\chi_t)\rangle\Big] + \frac{2}{\hbar}\mathbf{Im}\Big[\langle\,\hat{\mathbb{P}}^{KvN}(\chi_t)|\alpha_H(\chi_t)\rangle\Big]\\
&+\frac{1}{\hbar}\mathbf{Im}\Big[\langle\hat{\mathbf{P}}(\chi_t)|\hat{L}_H(\chi_t)\rangle\Big] + \frac{1}{\hbar}\mathbf{Im}\Big[\langle\,\hat{\mathbf{P}}(\chi_t)|\alpha_H(\chi_t)\rangle\Big].
\end{aligned}$$

**Remark 4.4.** *In the standard formalism, the dynamics would have consisted only in the first term. The red terms are the new contributions appearing in the Koopman-van Hove framework.*

### 4.3.4 Underlying group action

To get a deeper geometrical understanding of the situations where the average of the Koopman-van Hove linear momentum operator $\langle\hat{\mathbb{P}}\rangle$ is conserved, we will, similarly as the case of $\langle\hat{\mathbb{L}}\rangle$, determine a group action on $\mathcal{H}_{vH}$ such that $\langle\hat{\mathbb{P}}\rangle$ is a momentum map.

First, remark that

$$\begin{aligned}\langle\hat{\mathbb{P}}_\xi\rangle\,(\chi)(z) &= \left[-i\hbar\int\chi^*(z)\frac{\partial\chi}{\partial q}dz + \frac{1}{2}\int p\chi^*(z)\chi(z)dz\right]\cdot\xi\\ &= \hbar\int\chi^*(z)\Big(-i\frac{\partial\chi}{\partial q}+\frac{i}{2\hbar}p\Big)\cdot\xi\chi(z)dz\\ &= \hbar\int\chi^*(z)i\Big[\big(-\frac{\partial}{\partial q}(\cdot)+\frac{1}{2\hbar}p\cdot\mathbf{Id}\big)\cdot\xi\Big]\chi(z)dz\\ &= \hbar\,\langle\chi|i\hat{\mathbf{B}}_\xi\chi\rangle \qquad \text{with } \hat{\mathbf{B}}_\xi := \big(-\frac{\partial}{\partial q}(\cdot)+\frac{1}{2\hbar}p\cdot\mathbf{Id}\big)\cdot\xi. \end{aligned} \tag{4.17}$$

Then, by relation (4.10), the problem of determining a group action $\Phi$ generating $\langle\hat{\mathbb{P}}\rangle$ is equivalent to determining a group action $\Phi$ such that $\hat{\mathbf{B}}_\xi = \xi_{\mathcal{H}_{vH}} = \frac{d}{dt}\Big|_{t=0}\exp(t\xi)$ for all $\xi\in\mathbb{R}^3$. Hence, we have

**Theorem 4.5** (Group action generating $\langle\hat{\mathbb{P}}\rangle$ as a momentum map)**.** *Consider $G=\mathbb{R}^3$ and the action $\Phi:\mathbb{R}^3\times\mathcal{H}_{vH}\to\mathcal{H}_{vH}$ defined by*

$$\Phi(v,\chi)(z) := \Big(-\frac{\partial\chi}{\partial q}(z)+\frac{1}{2\hbar}p\chi(z)\Big)\cdot v. \tag{4.18}$$

*Then, $\langle\hat{\mathbb{P}}\rangle$ is the momentum map associated with the action $\Phi$ of $\mathbb{R}^3$ on $\mathcal{H}_{vH}$.*

*Proof.* Fix $\xi\in\mathfrak{g}\equiv\mathbb{R}^3$. Let $\{g_t\}_{t\in[a,b]}\subset\mathbb{R}^3$ be a 1-parameter family of group elements in $\mathbb{R}^3$ such that $g_0=\mathbf{Id}$ and $g_t'=\xi$.

By definition,

$$\begin{aligned}\xi_{\mathcal{H}_{vH}}(\chi)(z) &\overset{(2.11)}{=} \frac{d}{dt}\Big|_{t=0}\exp(t\xi)\chi(z)\\ &\overset{(4.18)}{=} \frac{d}{dt}\Big|_{t=0}\Big(-\frac{\partial\chi}{\partial q}(z)+\frac{1}{2\hbar}p\chi(z)\Big)\cdot g_t\end{aligned}$$

$$= \Big( -\frac{\partial \chi}{\partial q}(z) + \frac{1}{2\hbar} p\chi(z) \Big) \cdot \xi$$
$$\stackrel{!}{=} \hat{\mathbf{B}}_\xi(\chi)$$

$\overset{(4.10)}{\Longleftrightarrow}$ $\Phi$ is indeed the group action generating $\langle \hat{\mathbb{P}} \rangle$ as a momentum map. □

## 4.4 Dynamics in the Heisenberg picture

So far, the time evolution of the system we have been studying have only been encoded in the van Hove wave function. In quantum mechanics, this way of describing the dynamics is known as the Schrödinger picture, where the wave functions are allowed to have a time-dependence but the operators are not. In this section, we will introduce an alternative way of describing the time evolution, called the Heisenberg picture. We will also state an important result on the dynamics of Hermitian operators expressed within the Heisenberg picture.

### 4.4.1 The Heisenberg picture

Whereas the operators in the Schrödinger picture are not allowed to depend on time, their analogues in the Heisenberg picture are. Also, the converse holds for the wave functions: they can be time dependent in the Schrödinger picture but not in the Heisenberg one. In other words, the Heisenberg picture encodes the time dependence in the operators while the Schrödinger one incorporates it in the wave functions. More formally, displacing the time dependence of the wave functions to the operators is made as follows [23].

**Lemma 4.2** (Existence of propagators). *For every van Hove wave function $\chi_t \in \mathcal{H}_{vH}$ solution of the van Hove-Liouvillian equation of motion* (3.20)*, there exists a time-parametrized family $\{U_t\}_{t\in I} \subset U(6,\mathbb{R})$ in the group of unitary matrices and a van Hove wave function $\chi \in \mathcal{H}_{vH}$ such that*

$$\chi_t = U_t(\chi). \tag{4.19}$$

*Proof.* As the van Hove-Liouvillian equation of motion (3.20) is a linear equation with constant coefficients, its solution is unique by the Picard-Lindelöf theorem. Conclude by observing that $\chi_t = \exp[-\frac{i}{\hbar}\hat{\mathcal{L}}_H]\chi_0 := U_t(\chi)$ solves (3.20) and that $U_t \in U(6,\mathbb{R})$. □

**Remark 4.5.** *Note that the unitary operator $U_t$ depends only on $H$.*

**Definition 4.4** (Heisenberg picture of an operator)**.** *Consider an operator $A \in \mathrm{Herm}(\mathcal{H}_{vH})$ in the Schrödinger picture (time independant by definition). The **Heisenberg picture $A^{\mathrm{H}}$ of the operator $A$** is defined as $A^{\mathrm{H}} := \big(A \circ U_t\big)$.*

**Remark 4.6.** *The Heisenberg picture of an operator $A \in \mathrm{Herm}(\mathbb{R}^6)$ is well defined by lemma 4.2.*

### 4.4.2 Dynamics of $\hat{\mathbb{L}}$ and $\hat{\mathbb{P}}$ in the Heisenberg picture

Within the Heisenberg picture, the van Hove transformation $\hat{\mathcal{L}}$ establishes a strong link [1] between the dynamics of Hamiltonian functions $H \in C^\infty(\mathbb{R}^{2n})$ and their associated operators $\hat{\mathcal{L}}_H \in \mathrm{Herm}(\mathbb{R}^{2n})$.

**Theorem 4.6** (Observable dynamics in the Heisenberg picture)**.** *Let $A \in C^\infty(\mathbb{R}^{2n})$ be a Hamiltonian function. Then,*

$$\frac{d}{dt}\hat{\mathcal{L}}_A^{\mathrm{H}} = \frac{i}{\hbar}\big[\hat{\mathcal{L}}_H^{\mathrm{H}},\, \hat{\mathcal{L}}_A^{\mathrm{H}}\big] = \hat{\mathcal{L}}_{\{A,H\}}^{\mathrm{H}} \tag{4.20}$$

*Proof.* Let $\chi \in \mathcal{H}_{vH}$. Develop $\frac{d}{dt}\hat{\mathcal{L}}_A(\chi)$ using definition (3.20) and the chain rule (3.20). Remark that as $U_t \in U(2n,\mathbb{R})$, hence $\frac{d}{dt}U_t \in \mathfrak{u}(2n,\mathbb{R}) = \{M \in O(2n,\mathbb{R}) \mid M^\dagger = -M\}$. Conclude by regrouping the terms. □

In particular, for Hamiltonian functions $H$ evaluated along solutions $z$ of $\dot{z}(t) = X_H(z(t))$, the dynamics of $\hat{\mathbb{L}}$ and $\hat{\mathbb{P}}$ takes the following form.

**Corollary 4.3** (Dynamics of $\hat{\mathbb{L}}$ and $\hat{\mathbb{P}}$)**.** *Let $\xi \in \mathbb{R}^3$. Let $z \equiv (q,p) : I \to \mathbb{R}^6$ be a solution of $\dot{z}(t) = X_H(z(t))$. Then, for $J_\xi^1(q,p) = q \times p \cdot \xi$ and $J_\xi^2(q,p) = p \cdot \xi$ restricted to $z$, we have*

*(i)* $\displaystyle \frac{d}{dt}\hat{\mathbb{L}}_\xi^{\mathrm{H}} \overset{(4.20)}{=} \hat{\mathcal{L}}_{\{J_\xi^1,H\}}^{\mathrm{H}} = \hat{\mathcal{L}}_{q\times\dot{p}\cdot\xi}^{\mathrm{H}}$

*(ii)* $\displaystyle \frac{d}{dt}\hat{\mathbb{P}}_\xi^{\mathrm{H}} \overset{(4.20)}{=} \hat{\mathcal{L}}_{\{J_\xi^2,H\}}^{\mathrm{H}} = \hat{\mathcal{L}}_{\dot{p}\cdot\xi}^{\mathrm{H}}$

*Proof.* Remark that if $z : I \to \mathbb{R}^6$ is a solution of $\dot{z}(t) = X_H(z(t))$, then $\frac{d}{dt}F(z(t)) = \{F,H\}(z(t))$. Also, note that the restriction on z implies that $p = m\dot{q}$. Hence, $\big(\dot{q}\times p + q\times\dot{p}\big)\cdot\xi = q\times\dot{p}\cdot\xi$. □

**Corollary 4.4** (Newton's second law)**.** *Let $\xi \in \mathbb{R}^3$. Let $z \equiv (q,p) : I \to \mathbb{R}^6$ be a solution of Hamilton's equation $\dot{z}(t) = X_H(z(t))$. Then,*

$$\frac{d}{dt}\hat{\mathbb{P}}^{\mathrm{H}}_{\xi} \overset{(4.20)}{=} \hat{\mathcal{L}}^{\mathrm{H}}_{F(q)\cdot\xi} \tag{4.21}$$

*i.e. the variation of the van Hove linear momentum (projected on $\xi$) is the van Hove transformation of the total force $F$ (projected on $\xi$) acting on the system.*

*Proof.* As the motion is restricted on the one of a solution of $\dot{z}(t) = X_H(z(t))$, $p = m\dot{q}$ and $z$ satisfies Newton's law by theorem 2.1. Use corollary and the linearity of the van Hove transformation $\hat{\mathcal{L}}$. □

**Corollary 4.5** (Conservation of linear and angular momentum)**.** *Let $\xi \in \mathbb{R}^3$. Consider a physical system not acted by external forces. Then,*

*(i)* $\frac{d}{dt}\hat{\mathbb{P}}^{\mathrm{H}}_{\xi} \overset{(4.21)}{=} 0$

*(ii)* $\frac{d}{dt}\hat{\mathbb{L}}^{\mathrm{H}}_{\xi} \overset{(Cor. 4.3)}{=} 0$

*Proof.* For $\hat{\mathbb{L}}$, as the motion is restricted on the one of a solution of $\dot{z}(t) = X_H(z(t))$, $p = m\dot{q}$ and $z$ satisfies Newton's law by theorem 2.1. The result for $\hat{\mathbb{P}}$ is a consequence of corollary 4.4. □

**Corollary 4.6** (Ehrenfest equations)**.** *Let $\xi \in \mathbb{R}^3$. Let $z \equiv (q,p) : I \to \mathbb{R}^6$ be a solution of $\dot{z}(t) = X_H(z(t))$. Consider an Hermitian operator $\hat{A} \in \mathrm{Herm}(\mathcal{H}_{vH})$ where $\hat{A} = \hat{\mathcal{L}}_A$ for $A \in C^\infty(\mathbb{R}^6)$. Then,*

$$\frac{d}{dt}\langle \hat{A}^{\mathrm{H}} \rangle \overset{(4.20)}{=} \frac{i}{\hbar}\langle [\hat{\mathcal{L}}^{\mathrm{H}}_H, \hat{A}^{\mathrm{H}}] \rangle \tag{4.22}$$

*Proof.* This is direct by theorem 4.6. □

# Chapter 5

# The Koopman-van Hove formulation of the Kepler problem

The Koopman-van Hove classical mechanics is now equipped with the notions of energy, linear and angular momentum. In the last chapter, we saw that the modified formalism implied the existence of new terms in the dynamics of $\hat{\mathbb{P}}$ and $\hat{\mathbb{L}}$. Computing these extra terms for known physical systems might allow us to compare the Koopman-van Hove dynamics with the Koopman-von Neumann one and get a better insight and understanding of the physics behind them.

In this chapter, we start by presenting the Koopman-van Hove framework for the reduced Kepler problem. We then highlight the new contributions in the dynamics of the Koopman-van Hove linear and angular momentum operators by working in the Schrödinger picture. Finally, we investigate a more general physical interpretation of their dynamics by using the Heisenberg picture and conclude by stating a result on the average motion.

## 5.1 The Koopman-van Hove setting of the Kepler problem

Consider now the probabilistic reduced Kepler problem *(c.f. Appendix A)*, equipped with the Hamiltonian function $H_{red}(q,p) = E_{Kinetic}+E_{Potential} = \frac{p^2}{2\mu}-\frac{\lambda}{|q|}$ and whose dynamics is described by a probability density function $\rho : \mathbb{R}^6 \to [0,1]$

The transition to the Koopman-van Hove framework will be done by considering the van Hove classical wave functions $\chi_t \in \mathcal{H}_{vH}$ parametrized by t and satisfying the modified fundamental prescription.

By plugging the Hamiltonian ($A.4$) in (3.19), one finds the following expression for the van Hove-Liouvillian

$$\begin{aligned}
\hat{\mathcal{L}}_{H_{red}}\chi_t(z) := & i\hbar\Big\{\big(\frac{p^2}{2\mu}-\frac{\lambda}{|q|}\big),\chi_t\Big\}(z) - \left(\frac{1}{2}\begin{pmatrix} q \\ p \end{pmatrix} \cdot \begin{pmatrix} \frac{-\lambda q}{|q|^3} \\ \frac{p}{\mu} \end{pmatrix} - H(q,p)\right)\chi_t(z) \\
&= i\hbar\Big\{\big(\frac{p^2}{2\mu}-\frac{\lambda}{|q|}\big),\chi_t\Big\}(z) + \left(\frac{\lambda|q|^2}{2|q|^3} - \frac{p^2}{2\mu} + H(q,p)\right)\chi_t(z) \\
&= i\hbar\left(\frac{\lambda q}{|q|^3}\cdot\frac{\partial\chi_t}{\partial p} - \frac{p}{\mu}\cdot\frac{\partial\chi_t}{\partial q}\right)(z) - \frac{\lambda}{2|q|}\chi_t(z),
\end{aligned}$$

where $\mathbb{R}^3 \ni z := (p,q)$.

In particular, the **van Hove-Liouvillian equation of motion** for the probabilistic reduced Kepler problem is

$$i\hbar\frac{\partial}{\partial t}\chi_t(z) = i\hbar\left(\frac{\lambda q}{|q|^3}\cdot\frac{\partial\chi_t}{\partial p} - \frac{p}{\mu}\cdot\frac{\partial\chi_t}{\partial q}\right)(z) - \frac{\lambda}{2|q|}\chi_t(z). \tag{5.1}$$

## 5.2 Dynamics of $\hat{\mathbb{P}}$ and $\hat{\mathbb{L}}$

In the first place, we want to outline the new terms appearing in the Koopman-van Hove formalism by comparing them to the ones of the standard Koopman-von Neumann theory in the context of the Kepler problem and discuss their origins and contributions. Before doing this, note that

$$\begin{aligned}
\hat{\mathcal{L}}_{H_{red}}\chi_t(z) &= i\hbar\left(\frac{\lambda q}{|q|^3}\cdot\frac{\partial\chi_t}{\partial p} - \frac{p}{\mu}\cdot\frac{\partial\chi_t}{\partial q}\right)(z) - \frac{\lambda}{2|q|}\chi_t(z) \\
&= \hat{L}_{H_{red}}(\chi_t)(z) - \frac{\lambda}{2|q|}\chi_t(z),
\end{aligned} \tag{5.2}$$

*i.e.* the van Hove Liouvillian operator for the Kepler problem can be expressed as the standard Liouvillian operator plus the extra term $-\frac{\lambda}{2|q|}\chi_t(z)$.

**Remark 5.1.** *For the rest of this chapter, $H_{red} \equiv H$ and $\chi_t$ will denote a solution of the van Hove-Liouvillian equation of motion of the Kepler problem* (5.1).

## 5.3 Dynamics of $\hat{\mathbb{L}}$, $\langle\hat{\mathbb{L}}\rangle$, $\hat{\mathbb{P}}$ and $\langle\hat{\mathbb{P}}\rangle$

### Dynamics in the Schrödinger picture

Our goal here is to highlight the new contributions appearing in the Koopman-van Hove formalism. Then, using corollaries 4.1 and 4.2, equations (5.2) and (4.16), we obtain

$$\begin{aligned}
\frac{d}{dt}\hat{\mathbb{L}}(\chi_t)(z) &= -\frac{i}{\hbar}\big(\hat{\mathbb{L}}\circ\hat{L}_H\big)(\chi_t)(z) + \frac{\lambda i}{2\hbar|q|}\hat{\mathbb{L}}\chi_t(z),\\
\frac{d}{dt}\left\langle\hat{\mathbb{L}}\right\rangle(\chi_t) &= \frac{1}{\hbar}\Big(2\,\mathbf{Im}\big[\langle\hat{\mathbb{L}}(\chi_t)|\,\hat{L}_H(\chi_t)\rangle\big] - \lambda\,\mathbf{Im}\big[\langle\hat{\mathbb{L}}(\chi_t)|\,\tfrac{\chi_t}{|q|}\rangle\big]\Big),\\
\frac{d}{dt}\hat{\mathbb{P}}(\chi_t)(z) &= -\frac{i}{\hbar}\left(\hat{\mathbb{P}}^{\mathrm{KvN}}\circ\hat{L}_H\right)(\chi_t)(z) + \frac{\lambda}{2|q|}\Big(\frac{\partial}{\partial q}\chi_t(z) + \frac{q}{|q|^2}\chi_t(z)\Big) + \frac{i}{2\hbar}p\hat{L}_H(\chi_t) + \frac{\lambda i}{4\hbar|q|}p\cdot\chi_t(z),\\
\frac{d}{dt}\left\langle\hat{\mathbb{P}}\right\rangle(\chi_t) &= \frac{2}{\hbar}\mathbf{Im}\big[\langle\hat{\mathbb{P}}(\chi_t)|\,\hat{L}_H(\chi_t)\rangle\big] - \frac{\lambda}{\hbar}\mathbf{Im}\big[\langle\hat{\mathbb{P}}^{KvN}(\chi_t)|\,\tfrac{\chi_t}{|q|}\rangle\big]\\
&\quad + \frac{1}{\hbar}\mathbf{Im}\big[\langle\hat{\mathbf{P}}(\chi_t)|\,\hat{L}_H(\chi_t)\rangle\big] - \frac{\lambda}{2\hbar}\mathbf{Im}\big[\langle\hat{\mathbf{P}}(\chi_t)|\,\tfrac{\chi_t}{|q|}\rangle\big],
\end{aligned} \tag{5.3}$$

where the black terms denote the dynamics for the standard Koopman-von Neumann formalism and the colored ones the new contributions appearing in Koopman-van Hove one. *A full computation of the above expressions can be found in Appendix B.*

While the physical meaning of the above extra terms is not fully understood yet, some remarks can be made.

**Remark 5.2** (Different numbers of extra terms in $\hat{\mathbb{L}}$ and $\hat{\mathbb{P}}$)**.** *The new contributions for $\hat{\mathbb{L}}$ and $\langle\hat{\mathbb{L}}\rangle$ consist only in one term whereas the ones for $\hat{\mathbb{P}}$ and $\langle\hat{\mathbb{P}}\rangle$ consist in three. The reason for that originates from remark 4.1 and the fact that the Hamiltonian of the Kepler problem is not a quadratic Hamiltonian (c.f. theorem (3.2)).*

**Remark 5.3** (Origin of the contributions)**.** *The extra terms have three different origins: the red terms originate from the coupling of the standard Koopman-von Neumann operator with the extra term of the van Hove-Liouvillian, the blue terms originate from the coupling of the extra term of the Koopman-van Hove operator with the standard Liouvillian and finally, the green terms originate from the coupling between the extra terms of the Koopman-van Hove operator and the van Hove-Liouvillian.*

**Remark 5.4** (Behaviour for big $|q|$)**.** *All extra contributions become small for big $|q|$, expect the blue terms, namely the extra terms coming from the coupling of the extra term of the modified* *operator with the standard Liouvillian. The reason for that is that the extra term in the van Hove-Liouvillian (5.2) of the reduced Kepler problem is linear to $\frac{1}{|q|}$; as the blue term is the only whose not originating from it, it does not admit a $\frac{1}{|q|}$ dependance.*

## Dynamics in the Heisenberg picture

We saw that the previous expressions for the dynamics of $\hat{\mathbb{L}}$, $\langle\hat{\mathbb{L}}\rangle$, $\hat{\mathbb{P}}$ and $\langle\hat{\mathbb{P}}\rangle$ naturally highlighted the new contributions that did not figure in the Koopman-von Neumann formalism. However, understanding their physics in their current form is not obvious. In order to gain more insights on the dynamics of $\hat{\mathbb{L}}$ and $\langle\hat{\mathbb{P}}\rangle$, we will expose a more compact form of their dynamics by using the Heisenberg picture.

Let $\xi \in \mathbb{R}^3$. We restrict here the motion in $\mathbb{R}^{2n}$ to a solution $z : I \to \mathbb{R}^{2n}$ of $\dot{z} = X_H(z)$. Then, by using Newton's second law and by denoting the central force acting on the body by $F$, one finds

$$q \times \dot{p} \cdot \xi \overset{(A.1)}{=} \mu q \times F(q) \cdot \xi = 0, \tag{5.4}$$

where we used the fact that $p = \mu\, \dot{q}$ and that the force $F(q)$ is collinear with q.

Moreover, Newton's law allows us to write

$$\dot{p} \cdot \xi = \frac{1}{\mu} F(q) \cdot \xi \overset{(A.1)}{=} -\frac{\lambda}{\mu |q|^3} q \cdot \xi. \tag{5.5}$$

Equations (5.4), (5.5) and corollary 4.3 yield

$$\boxed{\begin{aligned} \frac{d}{dt}\hat{\mathbb{L}}^{\mathrm{H}}(\chi) &= 0, \\ \frac{d}{dt}\left\langle\hat{\mathbb{L}}^{\mathrm{H}}\right\rangle(\chi) &= 0, \\ \frac{d}{dt}\hat{\mathbb{P}}^{\mathrm{H}}(\chi)(z) &= \tfrac{1}{\mu}\hat{\mathcal{L}}^{\mathrm{H}}_{F(q)\cdot\xi} \overset{(3.20)}{=} \hat{L}^{\mathrm{H}}_{\dot{p}\cdot\xi} + \left(\dot{p}\cdot\xi - \tfrac{1}{2} q \cdot \tfrac{\partial}{\partial q}\tfrac{F(q)}{\mu}\right) \circ U_t \\ \frac{d}{dt}\left\langle\hat{\mathbb{P}}^{\mathrm{H}}\right\rangle(\chi) &\overset{(5.5)}{=} \tfrac{1}{\mu}\left\langle\chi \,\middle|\, \hat{\mathcal{L}}^{\mathrm{H}}_{F(q)\cdot\xi}\chi\right\rangle \end{aligned}} \tag{5.6}$$

Note that the first equation is equivalent to a fundamental conservation law.

**Theorem 5.1** (Conservation of the Koopman-van Hove angular momentum for the reduced Kepler problem). *Consider the Koopman-van Hove formulation of the reduced Kepler problem. Then, the Koopman-van Hove angular momentum operator* $\hat{\mathbb{L}}$ *(4.4) is constant.*

**Remark 5.5** (Interpretation of the dynamics of $\hat{\mathbb{P}}$ for the reduced Kepler problem). *In the Kepler problem, the dynamics of* $\hat{\mathbb{P}} \in \mathrm{Herm}(\mathcal{H}_{vH})$ *is untimely linked with the dynamics of* $\frac{1}{\mu}F(q)\cdot\xi = -\frac{\lambda}{\mu|q|^3}q\cdot\xi \in C^\infty(\mathbb{R}^{2n})$. *Note that this Hamiltonian function represents the projection of the force on a vector* $\xi \in \mathbb{R}^3$. *In particular, both dynamics are directly linked by the van Hove transformation in the Heisenberg picture as stated in theorem 4.6.*

We conclude this chapter by stating a property on the overall motion of the probabilistic reduced Kepler problem.

## 5.4 Planar motion

We are now interested to see here if the motion is planar as in the deterministic case. By definition,

$$\hat{\mathbb{L}}(\chi)(q,p)\cdot\hat{Q}\chi(q,p) = i\hbar\left(\frac{\partial\chi}{\partial q}\times q + \frac{\partial\chi}{\partial p}\times p\right)\cdot q\chi(z)$$
$$= i\hbar\frac{\partial\chi}{\partial p}\times p\cdot q\chi(z) = -i\hbar\, q\times p\cdot\left(\chi(z)\frac{\partial\chi}{\partial p}\right). \tag{5.7}$$

While $\hat{\mathbb{L}}\cdot\hat{Q}$ does not vanish, we will see that in average, the position operator is normal to the angular momentum operator.

**Theorem 5.2** (The average motion is on the plane). *Let* $\hat{\mathbb{L}}$ *be the Koopman-van Hove angular momentum operator and* $\hat{Q}$ *be the position operator defined as* $\hat{Q}\chi(q,p) = q\,\chi(q,p)$. *Then,* $\langle\hat{\mathbb{L}}\cdot\hat{Q}\rangle(\chi) = 0$ *for all* $\chi \in \mathcal{H}_{vH}$.

*Proof.* By definition,

$$\langle\hat{\mathbb{L}}\cdot\hat{Q}\rangle(\chi) = ih\int (p\times q)\cdot\frac{\partial\chi}{\partial p}\chi(z)\chi(z)^* dz = i\hbar\sum_{i=1}^{3}\int (p\times q)_i\frac{\partial\chi}{\partial p^i}\chi(z)\chi^*(z)dz$$

$$
\begin{aligned}
&= i\hbar \sum_{i=1}^{3} \lim_{R\to\infty} \left[ \cancel{\int (p\times q)_i \chi(z)^2 \chi^*(z) dq \Big|_{\partial B_R(0)\setminus\{0\}}} - \int_{B_R(0)\setminus\{0\}} \frac{\partial}{\partial p^i} \Big( (p\times q)_i \chi(z)\chi^*(z) \Big) \chi(z) dz \right] \\
&= -i\hbar \sum_{i=1}^{3} \int (p\times q)_i \cdot \left( \frac{\partial \chi}{\partial p^i}\chi^* + \chi \frac{\partial \chi^*}{\partial p^i} \right)(z)\, dz \\
&= -i\hbar \sum_{i=1}^{3} \left[ \int (p\times q)_i \frac{\partial \chi}{\partial p^i}(z)\chi^*(z) dz + \cancel{\lim_{R\to\infty} \int (p\times q)_i \chi(z)\chi^*(z) dq \Big|_{B_R(0)\setminus\{0\}}} \right. \\
&\qquad \left. - \int (p\times q)_i \frac{\partial \chi}{\partial p^i}(z)\chi^*(z) dz \right] \\
&= 0.
\end{aligned}
$$

□

**Remark 5.6** (Planar motion)**.** *As we showed that the van Hove angular momentum operator $\hat{\mathbb{L}}$ was constant (c.f. theorem 5.1), the result of theorem 5.2 implies that the average motion is on a plane.*

# Chapter 6

# The Koopman-van Hove formulation of the harmonic and anharmonic oscillators

Whether it is to describe a mass on a Hooke's Law spring, a particle moving near a local minimum of a potential or a lattice vibrating because of quantum fluctuations, the harmonic and anharmonic oscillators have shown to be simple common models which describe naturally many fundamentally complex systems in physics.

This chapter will be very similar to the previous one and will consist of 2 sections: the first one focuses on the harmonic oscillator and the second one on the anharmonic one. In both cases, and similarly to the last chapter, we expose their Koopman-van Hove formalism and investigate the dynamics of $\hat{\mathbb{L}}$ and $\hat{\mathbb{P}}$. In particular, we will see that the Koopman-van Hove dynamics for the harmonic oscillator is very similar to the Koopman-von Neumann one.

## 6.1 The harmonic oscillator

Without restriction of the generality, consider the 3-dimensional classical harmonic oscillator where the body have mass 1, spring constant $k$ and whose Hamiltonian $H$ is given by $H(q,p) := p^2 + \frac{kq^2}{2}$. By theorem 3.2, the Koopman-van Hove equation of motion remains unchanged.

### Dynamics in the Schrödinger picture

The dynamics of $\hat{\mathbb{L}}$, $\langle\hat{\mathbb{L}}\rangle$ , $\hat{\mathbb{P}}$ and $\langle\hat{\mathbb{P}}\rangle$ in the Koopman-van Hove framework satisfy

$$\boxed{\begin{aligned}
\tfrac{d}{dt}\hat{\mathbb{L}} &= -\tfrac{i}{\hbar}\big(\hat{\mathbb{L}}\circ\hat{L}_H\big)(\chi_t)(z),\\
\tfrac{d}{dt}\langle\hat{\mathbb{L}}\rangle &= \tfrac{2}{\hbar}\mathbf{Im}\big[\,\langle\hat{\mathbb{L}}(\chi_t)|\,\hat{L}_H(\chi_t)\rangle\,\big],\\
\tfrac{d}{dt}\hat{\mathbb{P}}(\chi_t)(z) &= -\tfrac{i}{\hbar}\big(\hat{\mathbb{P}}^{\mathrm{KvN}}\circ\hat{L}_H\big) - \tfrac{i}{2\hbar}\big(\hat{\mathbf{P}}\circ\hat{L}_H\big)(\chi_t)(z),\\
\tfrac{d}{dt}\hat{\mathbb{P}}(\chi_t)(z) &= \tfrac{2}{\hbar}\mathbf{Im}\big[\,\langle\hat{\mathbb{P}}^{\mathrm{KvN}}(\chi_t)|\,\hat{L}_H(\chi_t)\rangle\,\big] + \tfrac{1}{\hbar}\mathbf{Im}\big[\,\langle\hat{\mathbf{P}}(\chi_t)|\,\hat{L}_H\chi_t)\rangle\,\big].
\end{aligned}}$$

*A full computation of the above expressions can be found in Appendix B.*

**Remark 6.1** (Origin of the new contributions)**.** *As stated above, the fact that the Hamiltonian is quadratic implies, by theorem 3.2, that the Koopman-von Neumann and van Hove-Liouvillian operators (3.8) and (3.19) coincide. Then, as $\hat{\mathbb{L}}$ coincide as well with its standard analogue (defined using the same construction but with the standard Koopman-von Neumann transformation (3.8)), the dynamics of $\hat{\mathbb{L}}$ in the van Hove framework is completely similar to the one of the standard case. On the other side, the only new contributions come from the fact that $\hat{\mathbb{P}}$ does not coincide with $\hat{\mathbb{P}}^{KvN}$, and the coupling of this extra term with the Liouvillian operator $\hat{\mathcal{L}}_H = \hat{L}_H$ are highlighted in red.*

### Dynamics in the Heisenberg picture

More compact physical expressions can be derived by using the Heisenberg picture. First, remark that

$$\{q\times\dot{p}\cdot\xi, H\} = \{q\times F(q)\cdot\xi, H\} = 0, \tag{6.1}$$

$$\{\dot{p}\cdot\xi, H\} = -k\,q\cdot\xi. \tag{6.2}$$

Then, equations (6.1), (6.2) in theorem 4.6 yield

$$\boxed{\begin{aligned}
\tfrac{d}{dt}\hat{\mathbb{L}}^{\mathrm{H}}(\chi) &= 0,\\
\tfrac{d}{dt}\langle\hat{\mathbb{L}}^{\mathrm{H}}\rangle(\chi) &= 0,\\
\tfrac{d}{dt}\hat{\mathbb{P}}^{\mathrm{H}}(\chi)(z) &= -k\hat{\mathcal{L}}^{\mathrm{H}}_{q\cdot\xi} \overset{(3.20)}{=} -k\hat{L}^{\mathrm{H}}_{q\cdot\xi} - \tfrac{k}{2}(q\cdot\xi)U_t,\\
\tfrac{d}{dt}\langle\hat{\mathbb{P}}^{\mathrm{H}}\rangle(\chi) &= -k\,\langle\chi\,|\,\hat{\mathcal{L}}^{\mathrm{H}}_{q\cdot\xi}\chi\rangle\,.
\end{aligned}} \tag{6.3}$$

**Theorem 6.1** (Conservation of the Koopman-van Hove angular momentum for the reduced harmonic oscillator). *Consider the Koopman-van Hove formulation of the harmonic oscillator. Then, the Koopman-van Hove angular momentum operator $\hat{\mathbb{L}}$ (4.4) is constant.*

**Remark 6.2** (Interpretation of the dynamics of $\hat{\mathbb{P}}$ for the harmonic oscillator). *The dynamics of the projected van Hove linear momentum operator $\hat{\mathbb{P}}_\xi$ is described as the van Hove transformation of the projection $\dot{p} \cdot \xi$ of the classical linear momentum on $\xi$, or equivalently, as the van Hove transformation of the projection $\frac{1}{\mu}F(q) \cdot \xi = -k\, q \cdot \xi$ of the restauring force $F = -kq$ on $\xi$.*

## 6.2 The anharmonic oscillator

Without restriction of the generality, consider the 3-dimensional anharmonic oscillator where the body have mass 1 and spring constant $k$ and whose Hamiltonian $H$ is given by $H(q,p) := p^2 - a\,|q|^2 + b\,|q|^4$ for some $a, b \in \mathbb{R}$. In this case, the Hamiltonian is not quadratic. Equivalently, the van Hove-Liouvillian operator (3.19) will admit non-zero extra terms.

In this section, we will first compute the van Hove-Liouvillian operator for the anharmonic oscillator before using theorems 4.2 and 4.4 to expose the new Koopman-van Hove contributions to the dynamics of $\hat{\mathbb{P}}$ and $\hat{\mathbb{L}}$.

### 6.2.1 Determination of the van Hove-Liouvillian operator

By definition (3.19),

$$\begin{aligned} \hat{\mathcal{L}}_H &= \hat{L}_H - \left(\frac{1}{2}\hat{\mathbf{Z}} \cdot \nabla H - H\right) \\ &= i\hbar\left( - 2(a + 2b|q|^2)q\frac{\partial}{\partial q} - p\frac{\partial}{\partial p}\right) - b|q|^4\,\mathbf{Id}. \end{aligned} \tag{6.4}$$

In particular, the **van Hove-Liouvillian equation of motion** for the probabilistic anharmonic oscillator is

$$i\hbar\frac{\partial}{\partial t}\chi_t(z) = -i\hbar\left(2(a + 2b|q|^2)q\frac{\partial \chi_t}{\partial q} - p\frac{\partial \chi_t}{\partial p}\right)(z) - b|q|^4\,\chi_t(z). \tag{6.5}$$

### 6.2.2 Dynamics of $\hat{\mathbb{P}}$ and $\hat{\mathbb{L}}$

#### Dynamics in the Schrödinger picture

Theorem 4.2 yields

$$\begin{aligned}
\tfrac{d}{dt}\hat{\mathbb{L}}(\chi_t)(z) &= -\tfrac{i}{\hbar}\big(\hat{\mathbb{L}}\circ\hat{L}_H\big)(\chi_t)(z) - \tfrac{ib|q|^4}{\hbar}\big[\tfrac{\partial\chi_t(z)}{\partial q}\times q + \tfrac{\partial\chi_t(z)}{\partial p}\times p\big],\\
\tfrac{d}{dt}\langle\hat{\mathbb{L}}\rangle(\chi_t) &= \tfrac{2}{\hbar}\mathbf{Im}\big[\langle\hat{\mathbb{L}}(\chi_t)|\,\hat{L}_H(\chi_t)\rangle\big] - \tfrac{2b}{\hbar}\mathbf{Im}\big[\langle\hat{\mathbb{L}}(\chi_t|\,|q|^4\chi_t)\rangle\big],\\
\tfrac{d}{dt}\hat{\mathbb{P}}(\chi_t)(z) &= -\tfrac{i}{\hbar}\big(\hat{\mathbb{P}}^{\mathrm{KvN}}\circ\hat{L}_H\big)(\chi_t)(z) - 4b|q|^2q\chi_t(z) - b|q|^4\tfrac{\partial\chi_t}{\partial q}\\
&\quad -\tfrac{i}{2\hbar}\big(\hat{\mathbf{P}}\circ\hat{L}_H\big)(\chi_t)(z) + \tfrac{b}{2\hbar}p|q|^4\chi_t(z),\\
\tfrac{d}{dt}\langle\hat{\mathbb{P}}\rangle(\chi_t) &= \tfrac{2}{\hbar}\big[\langle\hat{\mathbb{P}}^{\mathrm{KvN}}(\chi_t)|\,\hat{L}_H(\chi_t)\rangle\big] + 2b\,\mathbf{Im}\big[\langle i\tfrac{\partial}{\partial q}(\chi_t)|\,|q|^4\chi_t)\rangle\big]\\
&\quad +\tfrac{1}{\hbar}\mathbf{Im}\big[\langle\hat{\mathbf{P}}(\chi_t)|\hat{L}_H(\chi_t)\rangle\big] - \tfrac{b}{\hbar}\mathbf{Im}\big[\langle\,\hat{\mathbf{P}}(\chi_t)|\,|q|^4\chi_t\rangle\big],
\end{aligned}$$

where the black terms denote the dynamics for the standard Koopman-von Neumann formalism and the colored ones the new contributions appearing in the modified framework. *A full computation of the above expressions can be found in Appendix B.*

Again, the physical meaning of the above extra terms have not been totally understood yet. Similarly to the case of the reduced Kepler problem, a few remarks can be made.

**Remark 6.3** (Different numbers of extra terms in $\hat{\mathbb{L}}$ and $\hat{\mathbb{P}}$)**.** *The new contributions for $\hat{\mathbb{L}}$ and $\langle\hat{\mathbb{L}}\rangle$ consist only in one term whereas the ones for $\hat{\mathbb{P}}$ and $\langle\hat{\mathbb{P}}\rangle$ consist in three. The reason for that originates from remark 4.1 and the fact that the Hamiltonian of the anharmonic oscillator is not a quadratic Hamiltonian (c.f. theorem (3.2)).*

**Remark 6.4** (Origin of the contributions)**.** *The extra terms have three different origins: the red terms originate from the coupling of the standard Koopman-von Neumann operator with the extra term of the van Hove-Liouvillian, the blue terms originate from the coupling of the extra term of the modified operator with the standard Liouvillian and finally, the green terms originate from the coupling between the extra terms of the Koopman-van Hove operator and the van Hove-Liouvillian.*

**Remark 6.5** (Domination of the extra terms for large $|q|$)**.** *All extra contributions become big for big $|q|$, expect the blue terms, namely the extra terms coming from the coupling of the extra* *term of the modified operator with the standard Liouvillian. The reason for that is that the extra term in the van Hove-Liouvillian (5.2) of the reduced Kepler problem is linear to $\frac{1}{|q|}$; as the blue term is the only whose not originating from it, it does not admit a $|q|^4$ dependance.*

### Dynamics in the Heisenberg picture

Similarly to the previous examples, we will first compute two fundamental quantities before applying theorem 4.6. Then,

$$\{q \times \dot{p} \cdot \xi, H\} \overset{(2.4)}{=} \{q \times (2aq - 4|q|^2 q) \cdot \xi, H\} = 0, \tag{6.6}$$

$$\{\dot{p} \cdot \xi, H\} \overset{(2.4)}{=} \xi \cdot \big(2aq - 4|q|^2 q\big). \tag{6.7}$$

$$\boxed{\begin{aligned} \tfrac{d}{dt}\hat{\mathbb{L}}^{\mathrm{H}}(\chi) &= 0, \\ \tfrac{d}{dt}\left\langle \hat{\mathbb{L}}^{\mathrm{H}} \right\rangle(\chi) &= 0, \\ \tfrac{d}{dt}\hat{\mathbb{P}}^{\mathrm{H}}(\chi)(z) &= -\hat{\mathcal{L}}^{\mathrm{H}}\big(\xi \cdot (2aq - 4|q|^2 q)\big), \\ \tfrac{d}{dt}\left\langle \hat{\mathbb{P}}^{\mathrm{H}} \right\rangle(\chi) &= -k \left\langle \chi \,\middle|\, \hat{\mathcal{L}}^{\mathrm{H}}_{q\cdot\xi}\chi \right\rangle. \end{aligned}} \tag{6.8}$$

**Theorem 6.2** (Conservation of the Koopman-van Hove angular momentum for the anharmonic oscillator)**.** *Consider the Koopman-van Hove formulation of the anharmonic oscillator. Then, the Koopman-van Hove angular momentum operator $\hat{\mathbb{L}}$ (4.4) is constant.*

**Remark 6.6** (Interpretation of the dynamics of $\hat{\mathbb{P}}$ for the anharmonic oscillator)**.** *The dynamics of the projected van Hove linear momentum operator $\hat{\mathbb{P}}_\xi$ is described as the van Hove transformation of the projection $\dot{p} \cdot \xi$ of the classical linear momentum on $\xi$, or equivalently, as the van Hove transformation of the projection $\frac{1}{\mu}F(q) \cdot \xi = \xi \cdot \big(2aq - 4|q|^2 q\big)$ of the overall force $F(q) = \mu\big(2a - 4|q|^2\big)q$ on $\xi$.*

# Chapter 7

# Conclusion and further work

In this report, we have exposed the standard Koopman-von Neumann formalism and its physical limitations. We then exposed a recent formalism, namely the Koopman-van Hove one [1], which solved these problems. We presented a natural construction of the linear and angular momentum operators and described their associated dynamics as well as that of their averages. We also determined the group action on $\mathcal{H}_{vH}$ generating their averages as momentum maps. We then applied these new objects to the Kepler problem, the harmonic, and anharmonic oscillators, isolated the new contributions coming from this new framework and related the dynamics of the Koopman-van Hove operators which we had constructed to the dynamics of Hamiltonian functions.

The Koopman-van Hove formulation of classical mechanics is promising for various reasons. First, as it corrects the physical limitations of the standard Koopman-von Neumann formulation, the use of geometrical mechanical tools allows a better understanding of the symmetries associated with physical observables. Also, because of the Lie algebra isomorphism $\hat{\mathcal{L}}$, the Lie algebra of the Hermitian operators is structurally the same as the one of the Hamiltonian functions; this has the advantage to offer a very strong structural link between the deterministic and probabilistic classical mechanics, and by extension with quantum mechanics.

On the other side, this new formalism comes with some complications when compared to the standard one. For example, the Liouvillian operator (3.19) and the van Hove linear momentum operator (4.15) admit in many cases extra non-trivial terms that make the dynamics more complex and the energy harder to compute. In particular, we have seen that they effect the dynamics of $\hat{\mathbb{L}}$ and $\hat{\mathbb{P}}$ in a non-negligible manner and that the physical interpretations of these contribution remains unclear, even for systems described by quadratic Hamiltonians (for example, the harmonic oscillator).

To finalize, we propose some future research directions and open questions that have emerged during the course of this work:

- Despite their mathematical origin being known, the three families of extra contributions for the dynamics of $\hat{\mathbb{L}}$ and $\hat{\mathbb{P}}$ are not physically understood. To answer this question, it might be helpful to start by investigating the fundamental physical meaning as well as the underlying group actions of the extra term $\mathbf{div}(\mathcal{J})$ appearing in the modified prescription (3.24). In addition to that, we gave a physical interpretation in terms of the force for the averages of $\hat{\mathbb{P}}$ and $\hat{\mathbb{L}}$ but their computations could be improved in order to recover their well known classical mechanics results.

- Also, we have been interested in describing the Kepler problem in this new framework but no mention to Kepler's laws have been made in this report. Even though the Koopman-van Hove is not a pathwise theory, it would be very interesting to investigate a form of their probabilistic analogues and derive them from fundamental principles. Moreover, the same applies to the Runge-Lenz and Hamilton's vectors which play a fundamental role in the understanding of symmetries of the deterministic Kepler problem [15, 7]. Constructing their analogues in a similar fashion to the way we did for $\hat{\mathbb{L}}$ and $\hat{\mathbb{P}}$ may lead to some understanding of a specific class of symmetries in probabilistic classical mechanics. Analogously to the deterministic case, the components of these new operators may as well satisfy relationships between them and give information about the shape and orientation of the overall motion [15].

- Recall figure 3.4.2. Consider a Hamiltonian $H \in C^\infty(M)$ for some manifold $M$. Knowing that the Liouville equation associated with this system can become very complicated to solve and is in general non-linear, its resolution might be facilitated by reformulating it in terms of a linear equation (for example a Liouvillian equation of motion by using a Koopman-von Neumann transformation). Now that a modified theory has been proposed, it would be interesting to explore the links between the solutions of both the standard and van Hove-Liouvillian equation of motion first with themselves and then the modified ones with the ones of the Liouville equation. Moreover, as $\hat{\mathcal{L}_H}$ is a bijection (c.f. remark 3.7), it is possible that the eigenvalues and eigenfunctions of the van Hove-Liouvillian operator serve as criteria on some properties of the solutions of the Liouville equation.

- Lastly, we have seen that the standard Koopman-von Neumann formalism has the same fundamental prescription as quantum mechanics, and because of their strong similarities, standard techniques of quantum mechanics have naturally been extended to the Koopman-von Neumann mechanics, for example perturbation methods, functional integration or diagram techniques. Since the prescriptions and the transformations in the Koopman-van Hove are different, one would naturally ask oneself if the above mentioned modifications have repercussions on the applications of these standard tools.

**Acknowledgements**

Thank you to my supervisor, Prof. D. Holm, for sharing with me his knowledge, experience, insight and love for geometric mechanics.

I would also want to express my profond gratitude to Dr. Tronci for his time in answering my questions on his work and for his sharp and insightful remarks.

And finally, thank you to all my beloved family and all my friends who have been part of the journey, in particular So, Daniela and Alvaro, for their support, valuable remarks concerning the format and help in the proof-reading.

# Appendix A

# The deterministic and probabilistic Kepler problem

By using Newton's second laws, the classical 2-body Kepler problem consists in a system of 2 ordinary differential equations for the position of respectively body A and B. However, a transformation of referential allows us to reformulate the problem as a 1-body problem with reduced mass. This will be presented in the following subsections.

## A.1 The classical model

Consider 2 particles A and B in $\mathbb{R}^3$ respectively with mass $m_A$ and $m_B$ at positions $r_A$ respectively $r_B$ attracting each other by a radial force $\mathbf{F} \equiv \mathbf{F}(r; r_0)$ centered in $r_0 \in \mathbb{R}^3$ of the form

$$\mathbf{F}(r; r_0) := -\frac{\lambda}{|r - r_0|^3} \cdot (r - r_0) \qquad \text{where } \lambda \in (0; \infty). \tag{A.1}$$

Morever, particle A's force on B (whose vector representation is pointing from B to A) $\mathbf{F}$ will be denoted as $\mathbf{F}_{BA}$ and similarly for $\mathbf{F}_{AB}$,

$$\mathbf{F}_{AB}(r) := \mathbf{F}(r; r_B),$$

$$\mathbf{F}_{BA}(r) := \mathbf{F}(r; r_A).$$

Remark that

$$\mathbf{F}_{AB}(r_A) = -\mathbf{F}_{BA}(r_B). \tag{A.2}$$

### A.1.1 Reduction to a 1 body problem with reduced mass

Newton's equations of motion for the 2 bodies A and B are

$$\begin{cases} \mathbf{F}_{AB}(r_A) = m_A \cdot r_A \\ \mathbf{F}_{BA}(r_B) = m_B \cdot r_B \end{cases} \iff \begin{cases} \frac{1}{m_A}\mathbf{F}_{AB}(r_A) = r_A \\ \frac{1}{m_B}\mathbf{F}_{BA}(r_B) \overset{(A.2)}{=} -\frac{1}{m_B}\mathbf{F}_{AB}(r_A) = r_B \end{cases} \tag{Newton's equations}$$

To transform this 2 body problem into a 1 body problem, substract the second line of Newton's equations to the first one. We obtain

$$\begin{aligned} (\frac{1}{m_A} + \frac{1}{m_B})\mathbf{F}_{AB}(r_A) = r_A - r_B =: q \in \mathbb{R}^3 \setminus \{0\} \\ \iff \mu q = \mathbf{F}_{AB}(r_A) = \mathbf{F}(r_A - r_B; 0)) = \mathbf{F}(q, 0) \\ \iff \boxed{\mu q = \mathbf{F}(q, 0)}, \end{aligned} \tag{A.3}$$

where $\mu := (\frac{1}{m_A} + \frac{1}{m_B})^{-1}$ is the reduced mass.

To sum up, the Newton's equations of the 2 body in the Kepler problem are equivalent to those of a body of reduced mass $\mu$ moving at position $q$ experiencing a central force $F_{AB}(q)$ directed toward the origin $(0, 0)$.

Hence, the Hamiltonian $H_{red}$ for the reduced body is

$$H_{red}(q, p) = E_{Kinetic} + E_{Potential} = \frac{p^2}{2\mu} - \frac{\lambda}{|q|}. \tag{A.4}$$

## A.2 Probabilistic reduced body model

Next, the generalising transition from deterministic to probabilistic is done by changing the way the motion is described. Instead of studying the deterministic path $z(t) := (q(t), p(t)) \in T\mathbb{R}^3 \equiv$ $\mathbb{R}^6$, one rather considers the probability density $\rho : \mathbb{R}^3 \to [0;1]$ associated with the position and angular momentum of the body.

Moreover, using (*A*.4) in (3.1), one finds that $\rho$ evolves according to the following dynamic:

$$\begin{aligned} \frac{\partial \rho}{\partial t} &= \left\{ \frac{p^2}{2\mu} - \frac{\lambda}{|q|}, \rho \right\} \\ &= -\frac{\lambda q}{|q|^3} \frac{\partial \rho}{\partial p} - \frac{p}{\mu} \frac{\partial \rho}{\partial q}. \end{aligned} \tag{A.5}$$

# Appendix B

# Computations of the dynamics of $\hat{\mathbb{L}}$, $\langle\hat{\mathbb{L}}\rangle$, $\hat{\mathbb{P}}$ and $\langle\hat{\mathbb{P}}\rangle$

We show the computations of the results of the dynamics we presented in the Schrödinger picture

## B.1 The Kepler problem

Remark 4.1 stating that $\hat{\mathbb{L}}^{\mathrm{KvN}} := \hat{L}_{J_\xi} = \hat{\mathcal{L}}_{J_\xi} =: \hat{\mathbb{L}}$ for $J_\xi(q,p) := q \times p \cdot \xi$ tells us that the Koopman-von Neumann and Koopman-van Hove formalisms would yield the same operators by following the same construction we presented above. Nevertheless, when looking at their dynamics, one extra term appears.

### Extra contribution for the dynamics of $\hat{\mathbb{L}}$

By corollary 4.1 and (5.2), we have

$$\begin{aligned}
\frac{d}{dt}\hat{\mathbb{L}}(\chi_t)(z) &= -\frac{i}{\hbar}\left(\hat{\mathbb{L}} \circ \hat{L}_H\right)(\chi_t)(z) - \frac{\lambda}{2|q|}\left(\frac{\partial}{\partial q}\chi_t(z) \times q - \cancel{\chi_t(z) \cdot q \times q} + \frac{\partial}{\partial p}\chi_t(z) \times p\right) \\
&= -\frac{i}{\hbar}\left(\hat{\mathbb{L}} \circ \hat{L}_H\right)(\chi_t)(z) - \frac{\lambda}{2|q|}\left(\frac{\partial}{\partial q}\chi_t(z) \times q + \frac{\partial}{\partial p}\chi_t(z) \times p\right)s \\
&\overset{(4.4)}{=} -\frac{i}{\hbar}\left(\hat{\mathbb{L}} \circ \hat{L}_H\right)(\chi_t)(z) + \frac{\lambda i}{2\hbar|q|}\hat{\mathbb{L}}\chi_t(z).
\end{aligned} \tag{B.1}$$

### Extra contribution for the dynamics of $\langle\hat{\mathbb{L}}\rangle$

Similarly, we have

$$\begin{aligned}\frac{d}{dt}\left\langle \hat{L}\right\rangle(\chi_t) &= \frac{2}{\hbar}\mathbf{Im}\left[\langle\hat{\mathbb{L}}(\chi_t)|\,\hat{\mathcal{L}}_H(\chi_t)\rangle\right]\\ &\overset{(5.2)}{=} \frac{2}{\hbar}\left\{\mathbf{Im}\left[\langle\hat{\mathbb{L}}(\chi_t)|\,\hat{L}_H(\chi_t)\rangle\right]-\mathbf{Im}\left[\langle\hat{\mathbb{L}}(\chi_t)|\,\frac{\lambda}{2|q|}\chi_t)\rangle\right]\right\}\\ &= \frac{1}{\hbar}\left(2\,\mathbf{Im}\left[\langle\hat{\mathbb{L}}(\chi_t)|\,\hat{L}_H(\chi_t)\rangle\right]-\lambda\,\mathbf{Im}\left[\langle\hat{\mathbb{L}}(\chi_t)|\,\frac{\chi_t}{|q|}\rangle\right]\right).\end{aligned} \tag{B.2}$$

## B.2 Dynamics of $\hat{\mathbb{P}}$ and $\langle\hat{\mathbb{P}}\rangle$

Concerning $\hat{\mathbb{P}}$, we have seen that $\hat{\mathbb{P}}^{\mathrm{KvN}}(\chi)(z) := -i\hbar\frac{\partial\chi}{\partial q}(z) \neq -i\hbar\frac{\partial\chi}{\partial q}(z)+\frac{1}{2}p\,\chi(z) =: \hat{\mathbb{P}}(\chi(z))$ for $J(q,p) := p\cdot\xi$, or equivalently, the above construction for the Koopman-von Neumann and Koopman-van Hove Liouvillian operator yields different operators. This extra term scales linearly with the linear momentum of the particle. We will now see what effect this term and the one exposed in (5.2) have on the dynamics of $\hat{\mathbb{P}}$ and $\langle\hat{\mathbb{P}}\rangle$.

### Extra contribution for the dynamics of $\hat{\mathbb{P}}$

By corollary 4.2 and by using the definition (4.16), we have

$$\begin{aligned}\frac{d}{dt}\hat{\mathbb{P}} &= -\left(\frac{\partial}{\partial q}+\frac{i}{2\hbar}p\right)\left(\hat{L}_H\chi_t(z)-\frac{\lambda}{2|q|}\chi_t(z)\right)\\ &\overset{(4.15)}{=} -\frac{i}{\hbar}\left(\hat{\mathbb{P}}^{\mathrm{KvN}}\circ\hat{L}_H\right)(\chi_t)(z)+\frac{\lambda i}{2\hbar}\hat{\mathbb{P}}^{\mathrm{KvN}}\left(\frac{\chi_t}{|q|}\right)(z)+\frac{i}{2\hbar}p\cdot\hat{L}_H(\chi_t)+\frac{\lambda i}{4\hbar|q|}p\cdot\chi_t(z).\end{aligned} \tag{B.3}$$

Reworking the first extra red term yields

$$\frac{\lambda i}{2\hbar}\hat{\mathbb{P}}^{\mathrm{KvN}}\left(\frac{\chi_t}{|q|}\right)(z) = \frac{\lambda}{2}\frac{\partial}{\partial q}\left(\frac{\chi_t}{|q|}\right) = \frac{\lambda}{2}\left(\frac{1}{|q|}\frac{\partial}{\partial q}\chi_t(z)-\frac{q}{|q|^3}\chi_t(z)\right) = \frac{\lambda}{2|q|}\left(\frac{\partial}{\partial q}\chi_t(z)-\frac{q}{|q|^2}\chi_t(z)\right). \tag{B.4}$$

Then, (*B*.4) in (*B*.3) gives

$$\frac{d}{dt}\hat{\mathbb{P}} = -\frac{i}{\hbar}\left(\hat{\mathbb{P}}^{\mathrm{KvN}}\circ\hat{L}_H\right)(\chi_t)(z)+\frac{\lambda}{2|q|}\left(\frac{\partial}{\partial q}\chi_t(z)+\frac{q}{|q|^2}\chi_t(z)\right)+\frac{i}{2\hbar}p\hat{L}_H(\chi_t)+\frac{\lambda i}{4\hbar|q|}p\cdot\chi_t(z). \tag{B.5}$$

**Extra contribution for the dynamics of $\langle \hat{\mathbb{P}} \rangle$**

Similarly,

$$\begin{aligned} \tfrac{d}{dt} \langle \hat{\mathbb{P}} \rangle (\chi_t) &= \tfrac{2}{\hbar}\mathbf{Im}\Big[ \langle \hat{\mathbb{P}}(\chi_t) |\, \hat{L}_H(\chi_t) \rangle \Big] - \tfrac{\lambda}{\hbar}\mathbf{Im}\Big[ \langle \hat{\mathbb{P}}^{KvN}(\chi_t) |\, \tfrac{\chi_t}{|q|} \rangle \Big] \\ &+ \tfrac{1}{\hbar}\mathbf{Im}\Big[ \langle \hat{\mathbf{P}}(\chi_t) |\, \hat{L}_H(\chi_t) \rangle \Big] - \tfrac{\lambda}{2\hbar}\mathbf{Im}\Big[ \langle \hat{\mathbf{P}}(\chi_t) |\, \tfrac{\chi_t}{|q|} \rangle \Big]. \end{aligned} \tag{B.6}$$

## B.3 The harmonic oscillator

By theorem 3.2, the Koopman-van Hove equation of motion remains unchanged, and so does the dynamics of $\hat{\mathbb{L}}$ as this operator coincide with its standard Koopman-von Neumann analogue. On the other hand, as $\hat{\mathbb{P}}^{\text{KvN}} \neq \hat{\mathbb{P}}$, there are still new contributions on this side. Indeed,

$$\begin{aligned} \frac{d}{dt}\hat{\mathbb{P}}(\chi_t)(z) &\overset{(4.4)}{=} -\frac{i}{\hbar}\Big( \hat{\mathbb{P}}^{\text{KvN}} \circ \hat{L}_H \Big)(\chi_t)(z) + \frac{1}{2}\Big( \hat{\mathbf{P}} \circ \hat{L}_H \Big)(\chi_t)(z) \\ &= \frac{d}{dt}\hat{\mathbb{P}}^{\text{KvN}}(\chi_t)(z) + \frac{1}{2}\Big( \hat{\mathbf{P}} \circ \hat{L}_H \Big)(\chi_t)(z) \\ \frac{d}{dt} \langle \hat{\mathbb{P}} \rangle (\chi_t) &\overset{(4.4)}{=} \frac{2}{\hbar}\mathbf{Im}\Big[ \langle \hat{\mathbb{P}}^{\text{KvN}}(\chi_t) |\, \hat{L}_H(\chi_t) \rangle + \langle \frac{1}{2}\hat{\mathbf{P}}(\chi_t) |\, \hat{L}_H(\chi_t) \rangle \Big] \\ &= \frac{d}{dt} \langle \hat{\mathbb{P}}^{\text{KvN}} \rangle (\chi_t) - \frac{i}{\hbar^2}\mathbf{Im}\Big[ \langle \hat{\mathbf{P}}(\chi_t) |\, L_H(\chi_t) \rangle \Big]. \end{aligned} \tag{B.7}$$

## B.4 The anharmonic oscillator

Again, corollary 4.1 and 4.2 together yields

$$\begin{aligned} \frac{d}{dt}\hat{\mathbb{L}}(\chi_t)(z) &= -\frac{i}{\hbar}\big(\hat{\mathbb{L}} \circ \hat{L}_H\big)(\chi_t)(z) - \frac{ib}{\hbar}\Big[\frac{\partial}{\partial q}\big(|q|^4\, \chi_t(z)\big) \times q + \frac{1}{|q|^4}\frac{\partial \chi_t(z)}{\partial p} \times p\Big] \\ &= -\frac{i}{\hbar}\big(\hat{\mathbb{L}} \circ \hat{L}_H\big)(\chi_t)(z) - \frac{ib}{\hbar}\Big[\Big(\cancel{4|q|^3 q\chi_t(z)} + |q|^4\frac{\partial \chi_t(z)}{\partial q}\Big) \times q + |q|^4\frac{\partial \chi_t(z)}{\partial p} \times p\Big] \\ &= -\frac{i}{\hbar}\big(\hat{\mathbb{L}} \circ \hat{L}_H\big)(\chi_t)(z) - \frac{ib|q|^4}{\hbar}\Big[\frac{\partial \chi_t(z)}{\partial q} \times q + \frac{\partial \chi_t(z)}{\partial p} \times p\Big], \\ \frac{d}{dt} \langle \hat{\mathbb{L}} \rangle (\chi_t) &= \frac{2}{\hbar}\mathbf{Im}\big[ \langle \hat{\mathbb{L}}(\chi_t) |\, \hat{L}_H(\chi_t) \rangle \big] - \frac{2b}{\hbar}\mathbf{Im}\big[ \langle \hat{\mathbb{L}}(\chi_t |\, |q|^4 \chi_t) \rangle \big], \\ \frac{d}{dt}\hat{\mathbb{P}}(\chi_t)(z) &= -\frac{i}{\hbar}\big(\hat{\mathbb{P}}^{\text{KvN}} \circ \hat{L}_H\big)(\chi_t)(z) - b\frac{\partial}{\partial q}\big(|q|^4 \chi_t\big)(z) \\ &\quad - \frac{i}{2\hbar}\big(\hat{\mathbf{P}} \circ \hat{L}_H\big)(\chi_t)(z) + \frac{b}{2\hbar}p|q|^4 \chi_t(z) \end{aligned}$$

$$
\begin{aligned}
&= -\frac{i}{\hbar}\big(\hat{\mathbb{P}}^{\mathrm{KvN}} \circ \hat{L}_H\big)(\chi_t)(z) - 4b|q|^2 q\chi_t(z) - b|q|^4\frac{\partial \chi_t}{\partial q} \\
&\quad - \frac{i}{2\hbar}\big(\hat{\mathbf{P}} \circ \hat{L}_H\big)(\chi_t)(z) + \frac{b}{2\hbar}p|q|^4\chi_t(z) \\
&= -\frac{i}{\hbar}\big(\hat{\mathbb{P}}^{\mathrm{KvN}} \circ \hat{L}_H\big)(\chi_t)(z) - b|q|^4\frac{\partial \chi_t}{\partial q} \\
&\quad - \frac{i}{2\hbar}\big(\hat{\mathbf{P}} \circ \hat{L}_H\big)(\chi_t)(z) + b\big(\frac{1}{2\hbar}|q|^2 p - 4q\big)|q|^2\chi_t(z),
\end{aligned}
$$

$$
\begin{aligned}
\frac{d}{dt}\langle \hat{\mathbb{P}} \rangle(\chi_t) &= \frac{2}{\hbar}\big[\,\langle \hat{\mathbb{P}}^{\mathrm{KvN}}(\chi_t)|\,\hat{L}_H(\chi_t)\rangle\,\big] + 2b\,\mathbf{Im}\Big[\langle i\frac{\partial}{\partial q}(\chi_t)|\,|q|^4\chi_t)\rangle\Big] \\
&\quad + \frac{1}{\hbar}\mathbf{Im}\Big[\langle \hat{\mathbf{P}}(\chi_t)|\hat{L}_H(\chi_t)\rangle\Big] - \frac{b}{\hbar}\mathbf{Im}\Big[\langle\,\hat{\mathbf{P}}(\chi_t)|\,|q|^4\chi_t\rangle\Big].
\end{aligned}
$$